\documentclass{jaa}
\usepackage{natbib}
\usepackage{graphicx,times}             
\usepackage{amssymb}
\usepackage{xcolor}
\usepackage{pdflscape}
\usepackage{longtable} 

\usepackage[pagebackref=true]{hyperref}

\def\teff{T_\mathrm{eff}}
\def\logg{\log g}
\def\tses{T_\mathrm{s}}
\def\tobs{T_\mathrm{o}}
\def\loggses{\log g_\mathrm{s}}
\def\loggobs{\log g_\mathrm{o}}
\def\mses{M_\mathrm{s}}
\def\mobs{M_\mathrm{o}}
\def\sun{\odot}
\def\zcs{ZZ Ceti}
\def\awmp{\mathit{WMP}}

\begin{document}\sloppy
% \linenumbers
%%paper title
%%For line breaks \\ can be used within title
\title{Seismic test of the mass-radius relationship of hydrogen-atmospheric white dwarf stars}

%%author names are separated by comma (,)
%%use \and before the last author name
%%use a * along with the number separated by comma
%% for the  author for correspondence
%%\textsuperscript{number} is used for affiliation
%%\affilOne, \affilTwo etc., upto \affilTwentyfive is possible
%%Please note the first letter after \affil is capitalised in the command
%%

\author{Tianqi Cang\textsuperscript{1}, Jiayi Zhang\textsuperscript{1}, Jian-Ning Fu\textsuperscript{1,2,*}, He Zhao\textsuperscript{3}, Weikai Zong\textsuperscript{1,2}}
\affilOne{\textsuperscript{1}School of Physics and Astronomy, Beijing Normal University, No.19, Xinjiekouwai St, Haidian District, Beijing, 100875, China.\\}
\affilTwo{\textsuperscript{2}Institute for Frontiers in Astronomy and Astrophysics, Beijing Normal University, Shahe, Manjing Road, Beijing 102206, China.\\}
\affilThree{\textsuperscript{3}Departamento de Fisica y Astronomia, Facultad de Ciencias Exactas, Universidad Andres Bello, Fernandez Concha 700, 8320000 Santiago, Chile.}

%%escape two column mode for title, affiliation and abstract
%%by giving \twocolumn command as shown

\twocolumn[{

\maketitle

%%include \corres to print the corresponding author Email id
\corres{jnfu@bnu.edu.cn}

%%include \msinfo for
%%manuscript information such as
%%received, revised and accepted dates
%%
\msinfo{1 January 2015}{1 January 2015}

%%abstract
\begin{abstract}
The pulsation of white dwarfs provides crucial information on stellar parameters for understanding the atmosphere and interior structure of these stars. In this study, we present a comprehensive statistical analysis of known ZZ Ceti stars from historical literature. Our dataset includes stellar parameters and oscillation properties from 339 samples, with 194 of them having undergone asteroseismological analysis. We investigated the empirical instability strip of ZZ Ceti stars and confirmed the linear relationship between temperature and weighted mean pulsation periods (WMP). We found that the WMP distribution is well-described with two groups of stars with peak values at $\sim254$ s and $\sim719$ s. Using seismic mass and trigonometrical radii derived from GAIA DR3 parallaxes, we tested the mass-radius relationship of white dwarfs through observational and seismic analysis of ZZ Cetis. They are generally larger than the theoretical values, with the discrepancy reaching up to $\sim15\%$ for massive stars with a mass estimated by seismology.
\end{abstract}

%%insert keywords separated by 3 hyphens using \keywords{words}

}]
%%close the twocolumn escape here

%%include \doinum{number}for the DOI number in the header
%%include \volnum{number} for the volume number in the header
%%include \year{yyyy} for  year of publication in the header
%%include \pgrange{num--num} page range of article in the header
%%include \artcitid{num} for the article citation id
%%include \lp to print last page of the article
%%include \setcounter{page}{pagenum} for the exact starting page of the article

\doinum{12.3456/s78910-011-012-3}
\artcitid{\#\#\#\#}
\volnum{000}
\year{0000}
\pgrange{1--}
\setcounter{page}{1}
\lp{1}

\section{Introduction}
A white dwarf (WD) represents the final evolutionary stage of a low- to intermediate-mass star. Such stars constitute the destiny of approximately 97\% of stars in the universe (\citealt{2003PASP..115..763C}). WDs predominantly comprise degenerate matter, leading to a well-defined relationship between their mass and radius, as initially formulated by (\citealt{1935MNRAS..95..207C}). In a WD, a larger mass results in a higher density of electrons, thereby increasing electron degenerate pressure and reducing the stellar radius. Understanding the mass-radius relationship (MRR) of white dwarfs is vital for comprehending their internal structure, the Chandrasekhar mass limit, and for utilizing Type Ia supernovae as standard candles in the study of the accelerating universe (\citealt{1998AJ....116.1009R,1999ApJ...517..565P}).

Obtaining accurate mass and radius measurements for an empirical MRR from observations is crucial to constrain theoretical models (\citealt{2005A&A...441..689A,2017MNRAS.465.2849T,2017ApJ...848...11B,2023arXiv230900239S}). While white dwarfs in binary systems can provide independent measurements of both radius and mass, only a limited number of these samples have achieved the level of accuracy necessary for a reliable orbital solution (\citealt{1997ApJ...488L..43S,2005MNRAS.362.1134B,2015ApJ...813..106B,2017MNRAS.470.4473P}). These samples need to be sufficient to cover a complete cooling sequence. Other observational MRR practices partially rely on WD model assumptions. The most widely used technique for estimating stellar radius employs parallax measurements and atmospheric parameters (\citealt{1997A&A...325.1055V}). Similarly, mass estimations are typically based on precise spectroscopic fits that highly depend on the atmospheric models of white dwarfs \citealt{2009ApJ...696.1755T}. Therefore, the MRR remains semi-empirical (\citealt{2017MNRAS.465.2849T,2017ApJ...848...11B}) and reliant on these stellar parameters, e.g., effective temperature and surface gravity, which would benefit from more precise constraints. Recent WD parameter studies with HST COS spectra (\citealt{2023arXiv230900239S})

During the cooling sequence of a hydrogen-dominated (DA) white dwarf, the star becomes an oscillating variable within a narrow effective temperature range of approximately 12,000 K, known as the instability strip (\citealt{2008PASP..120.1043F,2010A&ARv..18..471A, 2012A&A...539A..87V}). These so-called DAV or ZZ Ceti stars exhibit g-mode non-radial pulsations with periods ranging from 100 to 1500 seconds and can be analyzed through asteroseismology (\citealt{2019A&ARv..27....7C}). Since the pulsation modes are sensitive to the core composition, this method is a powerful tool for determining the mass with high precision (\citealt{2008ARA&A..46..157W}). Moreover, the possible temperature of ZZ Ceti stars contributes little to the radius variation, which is an additional constraint on testing the MRR.

The accuracy of seismological analysis depends on the number of independent pulsating modes. Due to the faint brightness and low amplitude of oscillations of WDs, the mode identification of ZZ Ceti stars is significantly enhanced by high-precision, continuous, long-term photometric monitoring (\citealt{2017ApJS..232...23H}). Constructing the MRR with ZZ Ceti requires complete coverage of the stellar mass, which benefits from a large set of samples with comprehensive seismological analysis. Until 2016, approximately 180 ZZ Ceti stars had been identified (\citealt{2016IBVS.6184....1B}), with few of them being massive ($>1M_\sun$). Recent astronomical surveys, e.g., SDSS (\citealt{2012ACP....12..207K}), LAMOST (\citealt{2017ApJ...847...34S}), and GAIA (\citealt{2020AJ....160..252V}), have accelerated the identification of new white dwarfs and ZZ Ceti candidates. The accumulation of photometric monitoring data from space missions, e.g., KEPLER (\citealt{2023ApJ...948...74H}), TESS (\citealt{2022MNRAS.511.1574R}), and ZTF (\citealt{2021ApJ...912..125G}), has roughly doubled the samples of \zcs, and provided new pulsation modes for known stars. These can benefit the definition of empirical instability strip and MRR and offer opportunities to test recent updates of theoretical models (\citealt{2019AAph...632A.119C, 2023MNRAS.524.5929C, 2021A&A...646A..30A}).

In this study, we systematically collect historical data on ZZ Ceti stars to construct a comprehensive dataset aimed at advancing the statistical understanding of these objects and testing the MMR of WD. Section \ref{sec:data} outlines our data collection strategy for DAVs and establishes a complete catalog of ZZ Ceti stars. Section \ref{sec:stat} presents the statistical analysis, including the empirical relationship between effective temperature and surface gravity or weighted mean pulsating periods. Section \ref{sec:mrr} explores the semi-empirical MRR derived from ZZ Ceti observations and contrasts these findings with theoretical models. Finally, Section \ref{sec:diss} concludes the work, compares it with other research, and discusses further ideas of the results.

\section{Data collection and complement}
\label{sec:data}
The targets collected for the analysis in this work focus on the confirmed \zcs, with a mass range of $0.3-1.3 M_\sun$. Although extremely low mass (ELM) WD ($M<0.3M_\sun$) with hydrogen-dominated atmosphere can present pulsating signals, we excluded them in our study for their potential particular evolution trace and overestimated luminosity because of the possible binary system \citealt{2013MNRAS.436.3573H,2015ASPC..493..217B}. The collection was initially based on the \zcs~list from \citet{2016IBVS.6184....1B}. We revisited the original literature in their table and extracted key observed stellar parameters (i.e., the effective temperature $\tobs$, the surface gravity $\loggobs$, and the mass of stars $\mobs$) from the modeling of the photometric or spectrometric observations, as well as the oscillation characteristics (i.e., periods \& amplitudes). Furthermore, we collected the seismological stellar parameters ($\tses$, $\loggses$, and $\mses$) if an asteroseismology analysis was performed for the target. We subsequently expanded the collection to more recent works. 

\begin{figure*}
    \centering
    \includegraphics[width=0.7\textwidth]{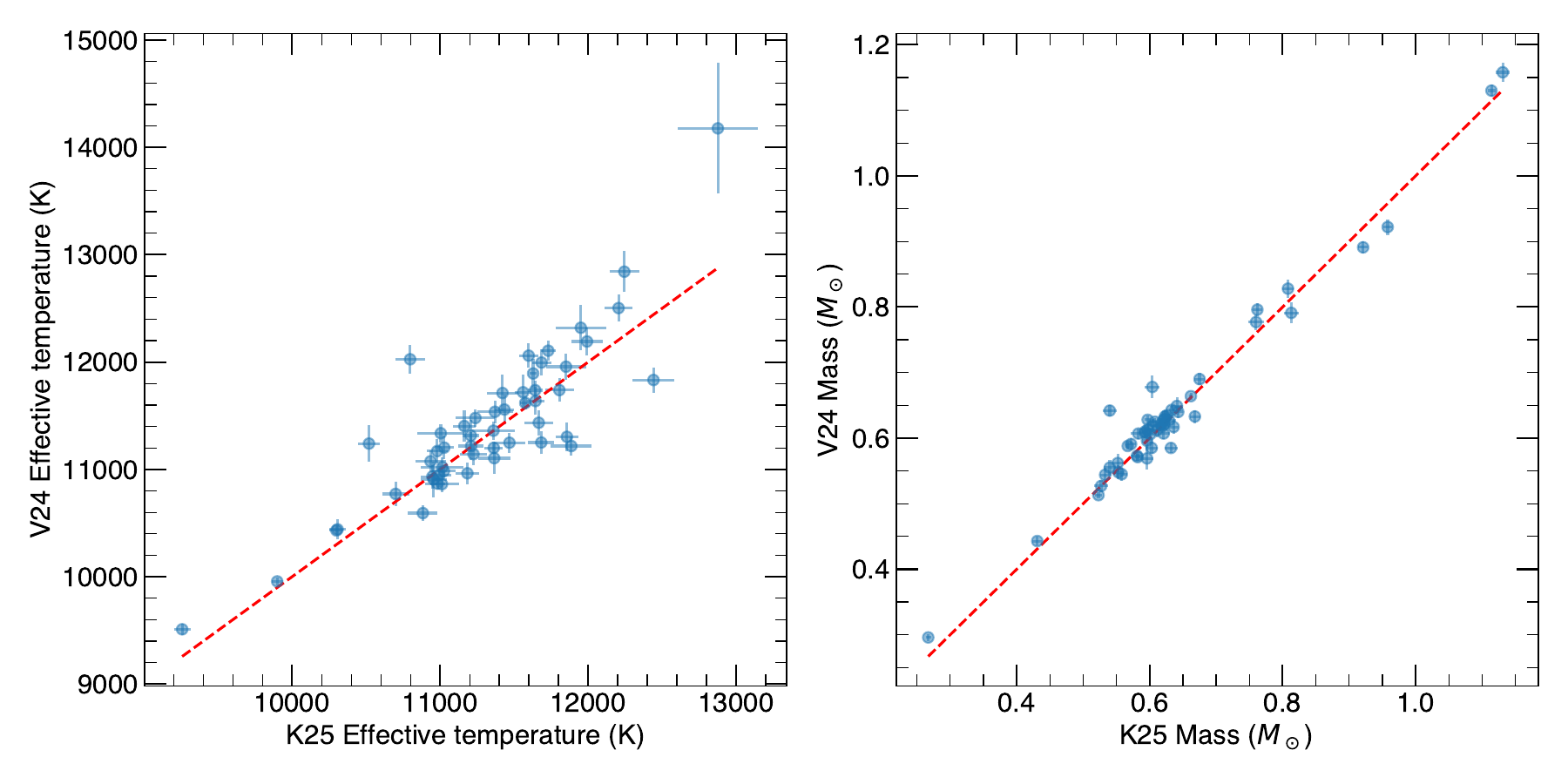}
    \caption{Comparison of observational effective temperature and mass between K25 and V24. The red dashed lines give the reference when the value from the two sources is equal.}
    \label{fig:K25_V24_par}
\end{figure*}

To merge information on the same target, we cross-matched our dataset with the GAIA DR3 database (\citealt{Vallenari2023}) using coordinates, which led to the identification of 339 independent \zcs. There is a subset of 194 \zcs~that underwent asteroseismology analysis, 167 of which used more than one oscillation mode. If there is more than one seismological modeling for a star in the same work, we kept the value with better model fitting, e.g., the quality function is given by \citet{2008MNRAS.385..430C}. For different works, we kept the parameters with less uncertainties of $\mses$. We excluded seismology results for stars with mass less than 0.5 $M_\sun$ from \citet{2022MNRAS.510..858R} since their models did not cover the low mass region. Missing $\loggobs$ or $\mobs$ in the source values were estimated by La Plata LPCODE evolutionary code \footnote{\url{http://evolgroup.fcaglp.unlp.edu.ar/TRACKS/tracks.html}} (\citealt{2005A&A...435..631A,2010ApJ...719..612A, 2012A&A...537A..33A,2012A&A...546A.119R, 2013A&A...555A..96S, 2016A&A...588A..25M, 2018MNRAS.473..457R}) with thick hydrogen layers ($M_H/M_*=10^{-4}$) for the pure-hydrogen interpolated, using the $\tobs$ and known $\logg$ or $\mobs$. Due to sensitivity to modeling choices, the mass of hydrogen ($\mathrm{M_H}$) is hard to compare and is excluded from our analysis. We cross-matched our \zcs~catalogue with three spectroscopic catalogues: the SDSS spectroscopic analysis for pure DA white dwarfs in 100 pc (\citealt{2025ApJ...979..157K}, hereafter K25), Gaia XP spectra analysis (\citealt{2024A&A...682A...5V}, hereafter V24), and SDSS spectra analysis from \citealt{2011ApJ...730..128T} (hereafter T11). Considering the accuracy and consistency of the data, we take the observational parameters ($\tobs$, $\loggobs$, and $\mobs$) in our analysis in the following order: K25, V24, T11, and others from the original works. Fig. \ref{fig:K25_V24_par} compares the $\teff$ and mass between K25 and V24. The $\teff$ from the two sources is generally consistent, and the mass measurements are well matched. Considering that the data from K25 and V24 occupy over 90\% of observational parameters, the statistical analysis based on these values is consistent.

Table \ref{tab:dav_info} lists stellar parameters of all collected DAVs. With the exception of WD J1216+0922, all \zcs~in our dataset have parallax measurements from GAIA and lie within 1,000 pc. Our dataset includes 9 targets confirmed to be in binary systems (labeled as "BS") in the original works and ten stars exhibiting infrared radiation excess (labeled as "IR") according to \citet{2020ApJ...902..127X} and \citet{2021ApJ...920..156L}. 

\section{Statistical characterization of \zcs~parameters}
\label{sec:stat}
\subsection{Empirical instability strip}
\label{sec:is} 
\begin{figure*}
    \centering
    \includegraphics[width=0.8\textwidth]{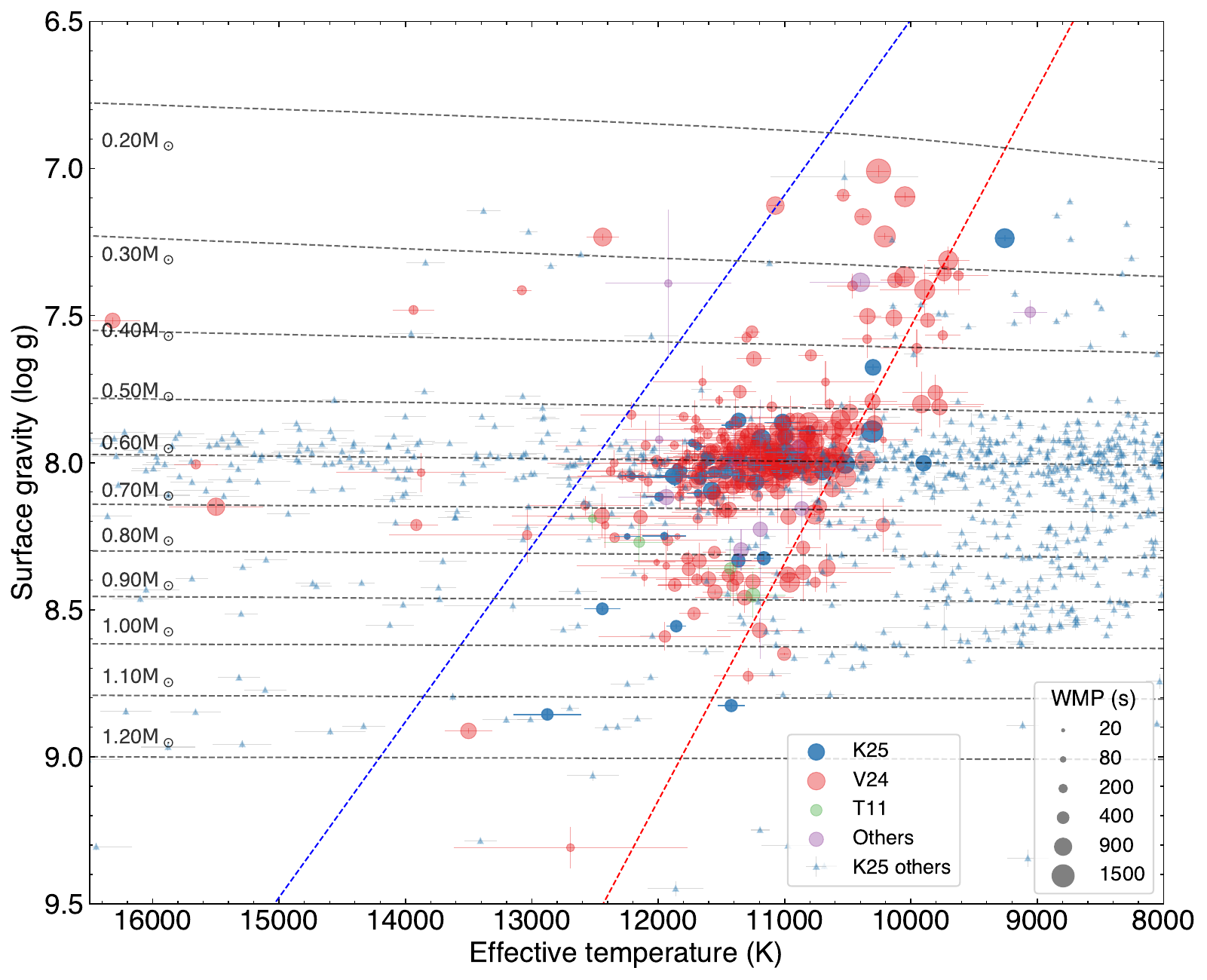}
    \caption{Distribution of ZZ Ceti stars on the $\teff$ - $\logg$ plane. Circles are known ZZ Ceti stars with parameters taken from \cite{2025ApJ...979..157K} (Blue), \cite{2024A&A...682A...5V} (Red), \cite{2011ApJ...730..128T} (Green), and values collected from the original paper (Purple), respectively. The size of the points shows the amplitude-weighted mean periods of each star. Small blue triangles are pure DA white dwarfs collected from \cite{2025ApJ...979..157K} except known DAVs. Blue and red dashed lines show the theoretical instability strip from \citet{2015ApJ...809..148T}. Horizontal gray dashed lines present the La Plate models of white dwarfs.}
    \label{fig:teff_logg}
\end{figure*}

Employing the observed effective temperature $\tobs$ and the surface gravity $\loggobs$ of these confirmed DAVs, we obtained an empirical instability strip in the $\teff-\logg$ diagram according to our data collection, as presented in Fig. \ref{fig:teff_logg}. Most of the known \zcs~locate inside the instability strip predicted by \citet{2015ApJ...809..148T} in $3\sigma$ of uncertainty, and the masses of them are grouped between 0.5-0.7 $M_\sun$. Samples located beyond the red edge (red dashed-line in Fig. \ref{fig:teff_logg}) of the instability strip exhibit a slight discrepancy at the edge ($<1000$K) across varying masses, suggesting that the observed red-edge of instability is cooler than the prediction. There are fewer samples near the blue edge, and the observational parameters can exhibit a large discrepancy from the blue edge. We check the three outliers (GAIA DR3 2260805780286092032, 1669095729418204416, and 5215511523500404352) with $\tobs>15000$, all of them are marked as a \zcs~in \citet{2022MNRAS.511.1574R} identified in the TESS observation. GAIA DR3 1669095729418204416 (TIC 232979174, $\tobs=$15658K, $\loggobs=$8.006) locates in the instability strip with parameters in \citet{2021MNRAS.508.3877G}. GAIA DR3 5215511523500404352 (TIC 804835539, $\tobs=$15499K, $\loggobs=$8.15) has an infrared excess, which may affect the determination of temperature. The seismological parameters of GAIA DR3 2260805780286092032 (TIC 229581336, $\tobs=$16317K, $\loggobs=$7.517) with three independent modes give a temperature of $\sim11000$K, suggesting that the model fitting to the GAIA XP spectra may be unsuccessful for this target. 

We also mark other pure DA white dwarfs in 100 pc, that are not identified as \zcs, in Fig. \ref{fig:teff_logg} with small blue triangles from \citet{2025ApJ...979..157K}. Over 100 nearby WDs can be found in the instability strip with no significant period detected (e.g., GD 344) or no timeseries photometric observation available.

\subsection{Amplitude-weighted mean period}
\begin{figure*}
    \centering
    \includegraphics[width=0.7\textwidth]{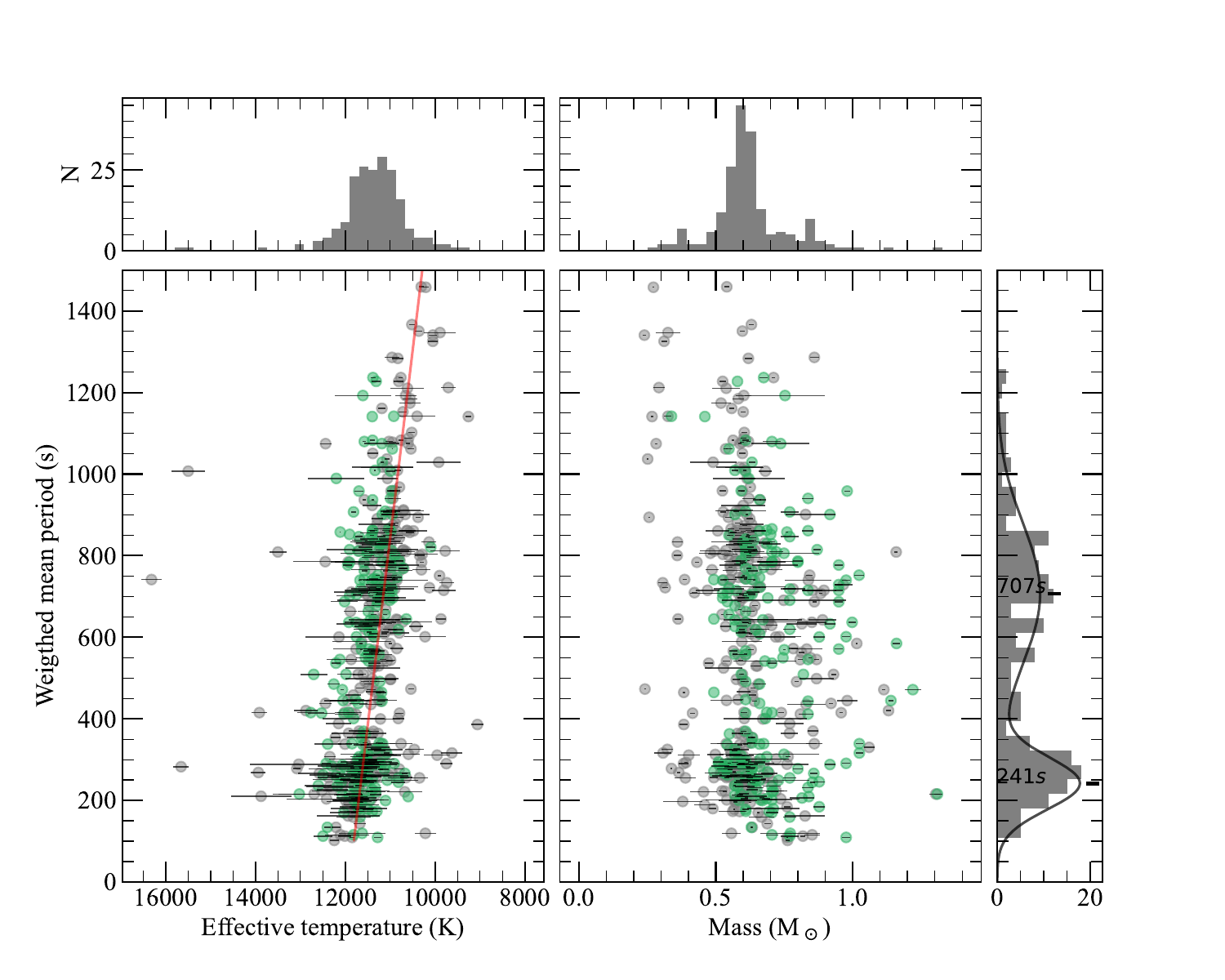}
    \caption{Mean weighted pulsating period (WMP) as a function of effective temperature (left) and stellar mass (right) for \zcs, as well as the distributions of observational parameters (grey). In comparison, green points show the seismology results. A two-gaussian model is used to fit the distribution of WMP, with peaks of $\sim241$s and $\sim707$s. The red line shows a linear relationship of $\awmp \propto -0.92\teff$, fitted by linear regression.}
    \label{fig:Teff_WMP}
\end{figure*}

The evolution of a WD is highly related to the cooling process, which also impacts the oscillation properties of the \zcs~(e.g., \citealt{2001PASP..113..409F}). For a cool \zcs, the excited period tends to be longer for the deeper convection zone, which was initially mentioned by \citet{1982ApJ...252L..65W} and discussed in detail in \citet{1992ApJS...81..747B}. To explore the relationship between the excited periods and the stellar parameters, a metric used to describe the pulsation properties of \zcs~is to calculate the amplitude-weighted mean pulsating period (WMP, \citealt{1993BaltA...2..407C,2008PASP..120.1043F, 2017ApJS..232...23H}), which can be used to cancel out amplitude effects due to different observation bands. We include all records of oscillation properties for the analysis and use a mean WMP for multi-epochs:
\begin{equation}
\label{eq:wmp}
    \mathrm{WMP} = \frac{1}{n}\sum^{n}_{i}\frac{\sum_{j} A_{j}P_{j}}{\sum_{j}A_{j}},
\end{equation}
where $P$ and $A$ are the detected period and the corresponding amplitude in a specific epoch. The size of markers in Fig. \ref{fig:teff_logg} represents the WMP of the targets.   We used the size of the circle in Fig.\ref{fig:teff_logg} to mark the $\awmp$ of the target, and the markers are generally larger in the top region compared to the bottom. Fig. \ref{fig:Teff_WMP} presents a diagram between $\tobs$ and $\awmp$. A linear fit can be used to describe the relationship between the two parameters:  
\begin{equation}
\label{eq:wmp_fit}
\mathrm{WMP}~\mathrm{(s)} = 10942(943) - 0.92(0.08)\times \teff~\mathrm{(K)}. 
\end{equation}
 We also explored the distribution between mass and WMP, but no significant trend was found. The sub-panel of Fig. \ref{fig:Teff_WMP} presented a histogram for WMP, and the distribution shows two main peaks of $241\pm64$s and $707\pm171$s.

\section{Empirical mass-radius relationship from \zcs}
\label{sec:mrr}
Asteroseismology analysis provides independent measurements for the stellar mass and $\teff$ of \zcs. The observational or trigonometrical radius $R_\pi$ of the star can be well estimated based on the astronomical distance measurements and corresponding apparent magnitude (\citealt{1997A&A...325.1055V, 2017MNRAS.465.2849T,2017ApJ...848...11B, 2021MNRAS.508.3877G}).

We employed the parallax from GAIA and the corresponding apparent G magnitude $m_G$, and derived the absolute Gaia G magnitude $M_G$ for each available target. The extinction correction was applied for further consideration. Since the WDs in our sample are very nearby (over 90\% are within 230\,pc) and the model of their intrinsic colors contains significant uncertainties, in this work we derived their monochromatic
extinctions at 541.4\,nm ($A_0$) by the recent released 3D dust map from \citet{Dharmawardena2024}\footnote{The dust map can be accessed online via \url{https://zenodo.org/records/11448780}.}. We then converted
$A_0$ to G--band extinction $A_{\rm G}$ considering the non-linear effects of both the stellar SED and the filter transmission (see detailed discussions in e.g., \citealt{Li2023,MaizApellaniz2024}). Over 90\% WDs
in our sample have $A_{\rm G}$ smaller than 0.194\,mag with a median of 0.084\,mag. 

\begin{figure*} 
    \centering
    \includegraphics[width=0.8\textwidth]{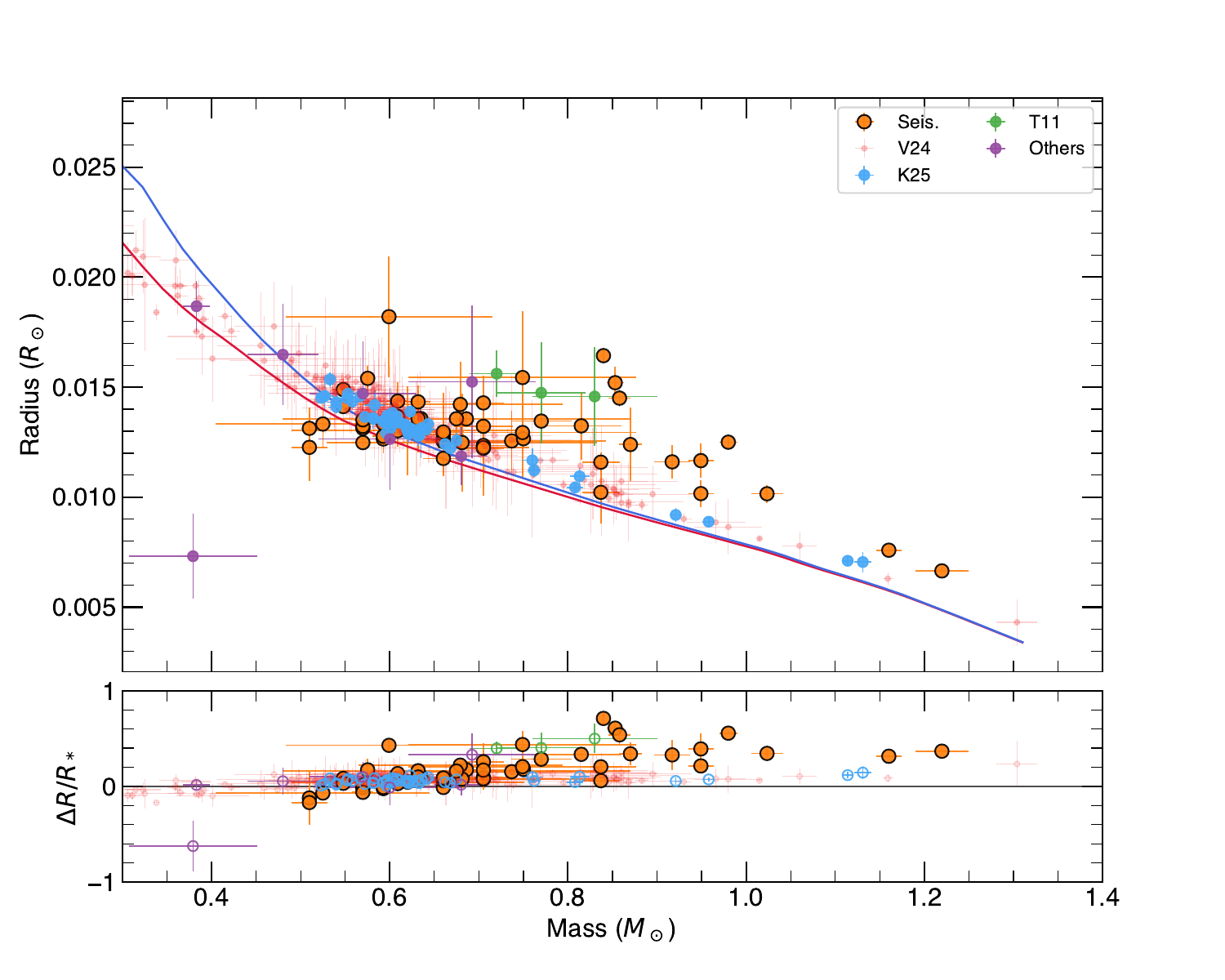}
    \caption{(Upper) Eperical mass-radius relationship (MRR) of \zcs. Radii are taken from trigonometrical radii $R_\pi$. Color points are derived from K25, V24, T11, and others, respectively, as described in Sect. \ref{sec:data}. Orange circles are derived from the seismology results, which include over six observed modes. Solid lines show the theoretical blue and red edge MRR of \zcs. (Lower) Residuals between $R_\pi$ and the theoretical radius as a fraction of $R_\pi$.}
    \label{fig:MRR}
\end{figure*}

The luminosity of the sample was represented by bolometric magnitude $M_{bolo} = M_G + BC$, where bolometric correction ($BC$) for Gaia G-band (\citealt{2022arXiv220605864C}) was derived from the $\tobs$ and $\loggobs$ with LPCODE model. The derived $BC$ is sensitive to $\teff$ and has a tiny dependence on the $\logg$ or models. Consider that the modeling of spectroscopic or photometric observation is sensitive to the temperature, as well as that the difference between $\tobs$ and $\tses$ is typically smaller than 1,000K (equivalent to $\Delta \mathrm{BC}\sim 0.1$), we used $\tobs$ in the estimation process of $BC$. Taking the absolute bolometric magnitude of the Sun as $M_{bolo,\sun}=+4.74$ (\citealt{2015arXiv151006262M}), we derived trigonometrical radii $R_\pi$ of the \zcs~by applying $L_*=4\pi R_\pi^2 \sigma \teff^4$ and $L_* / L_\sun = 10^{-0.4(M_\mathrm{bolo,*}-M_\mathrm{bolo,\sun})}$, where $\sigma$ is the Stefan-Boltzmann constant. Using $R_\pi$ and $\mobs$, we presented a semi-empirical MRR of \zcs~in  Fig. \ref{fig:MRR}. For stars with seismical analysis, we estimated the radii using the same method above and located the points with $\mobs$ and $R_\pi$ estimated by the $\tses$ in the figure. For comparison, we also included the theoretical MRR derived by LPCODE, with a thick ($M_H=10^{-4}$) atmosphere, and take an extended $\teff$ of 8000K and 15000K for red and blue edge.

\begin{figure*} 
    \centering
    \includegraphics[width=0.9\textwidth]{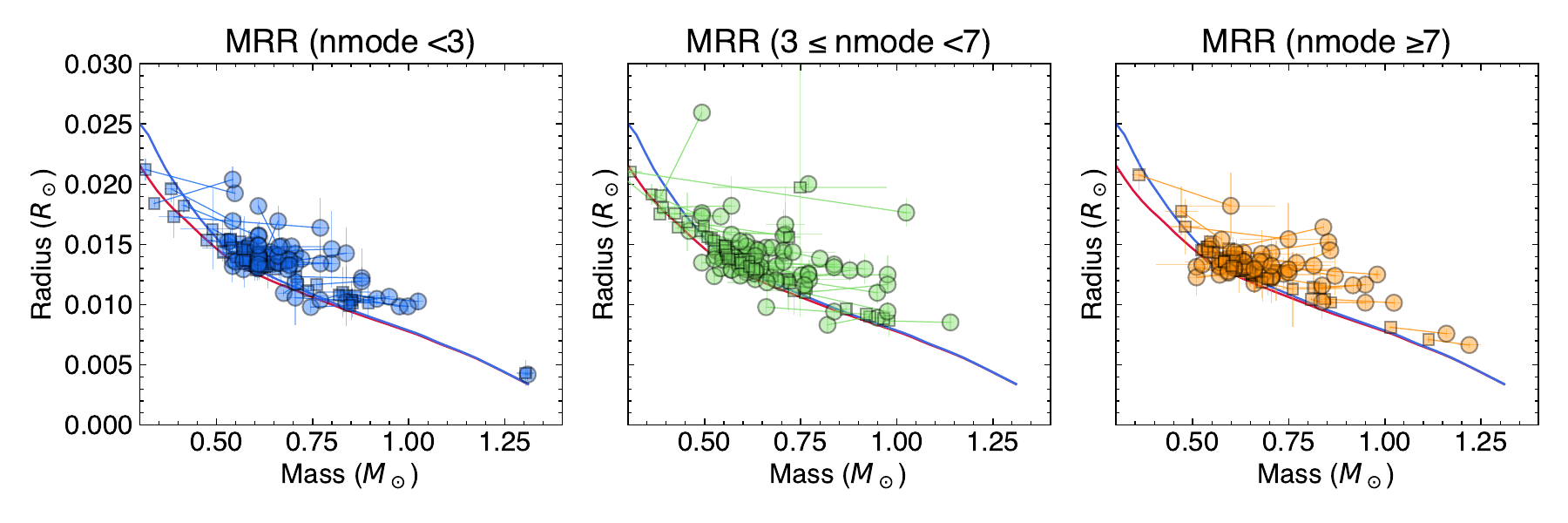}
    \caption{Comparison between observational (squares) and seismological (circles) results in MRR diagram. Three panels show the samples with $<3$, $3-7$, $>7$ observed modes, respectively. Observational and seismological results of one star are linked with solid lines.}
    \label{fig:MRR_compare}
\end{figure*}

\begin{figure*} 
    \centering
    \includegraphics[width=0.9\textwidth]{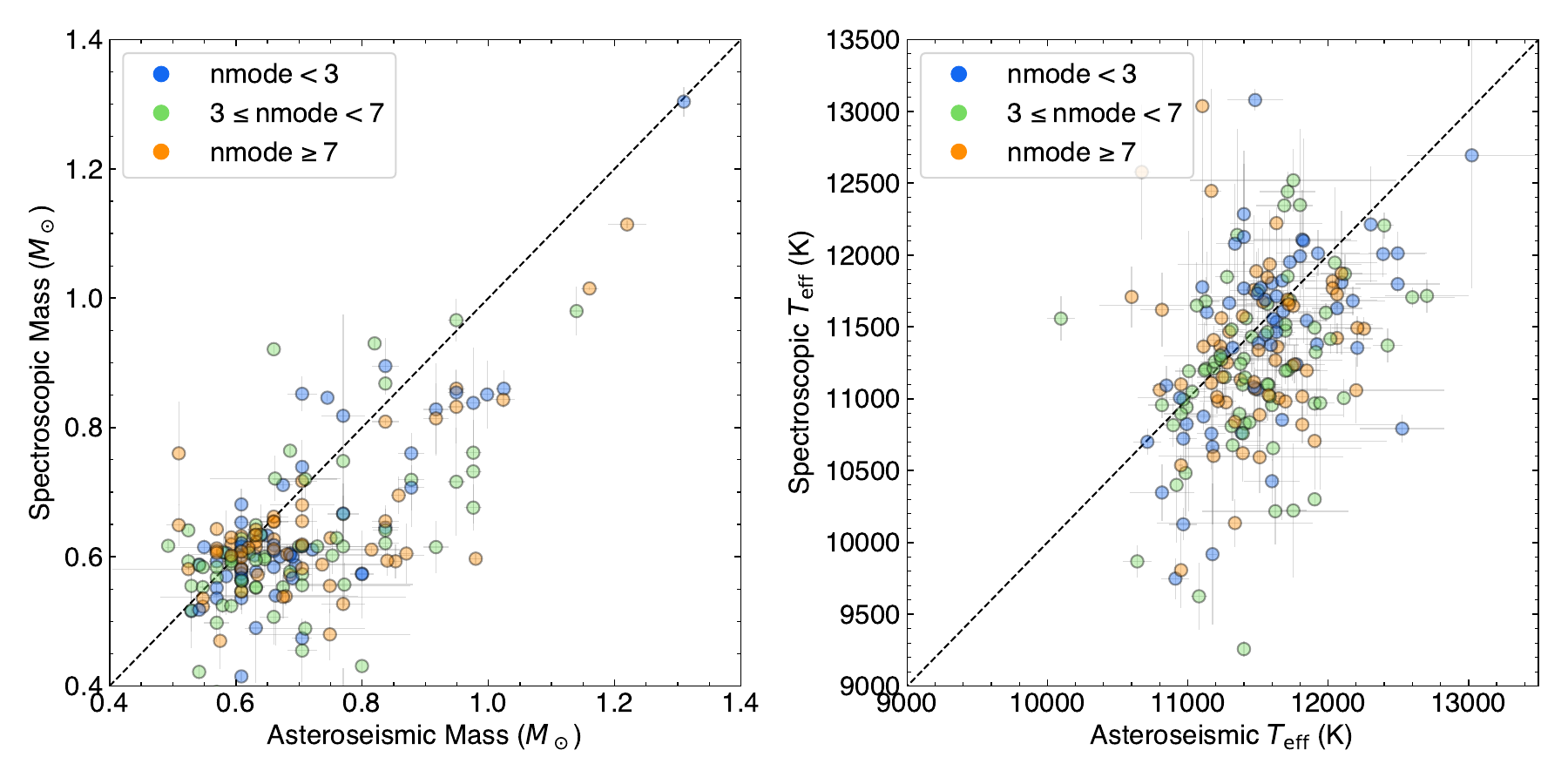}
    \caption{Comparison between observational and seismological results of \zcs. All observational measurements are derived from spectra. Modes used in the asteroseismology analysis are marked with different colors. (Left) Comparison between stellar mass (Right) and Comparison between effective temperature.}
    \label{fig:mass_compare}
\end{figure*}

In spite of a few extreme outliers, $R_\pi$ and $\mobs$ are generally consistent with theoretical predictions. However, when using seismically determined masses, we find that $R_\pi$ exceeds the theoretical values by 5–10\%, as shown in Fig.~\ref{fig:MRR}. This deviation becomes larger for more massive WD, particularly for $M>0.65\,M_\sun$. We compared the observational and seismological results for \zcs~in Fig.~\ref{fig:MRR_compare}, which shows that $R_\pi$ derived by the two methods remains consistent. As illustrated in Fig.~\ref{fig:mass_compare}, the dominant source of the discrepancy is the mass estimate, since the measured $R_\pi$ values agree within uncertainties. The number of detected modes does not significantly affect the offset, suggesting a systematic difference between seismic and spectroscopic mass determinations.

\section{Discussion and Conclusions}
\label{sec:diss}
In this work, we collected the most complete catalogue of \zcs~and collected the main stellar parameters and oscillation properties of them from the historical literature. 194 of 339 \zcs~were carried out with asteroseismology analysis. With the modeled $\teff$ and $\logg$ by the spectroscopy or photometry, we define a new empirical instability strip of \zcs. Most of the \zcs~in our collection are located inside the theoretical instability strip in $3\sigma$. Extreme blue outliers show a systematic dependency on the theoretical blue edge, which can be attributed to the unsuccessful measurements. However, blue outliers show a systematic shift to the blue edge of the instability strip, indicating that the red edge can extend to cooler WDs.  

We collected the observed oscillation properties of this \zcs~and calculated the WMP to present their typical oscillating periods. A clear linear relationship can be defined between WMP and $\teff$. The relationship suggests that a lower $\teff$ tends to have a longer period of pulsation, which is well predicted (e.g., \citealt{1982pccv.conf...46W}). The trend computed in our work is generally consistent with the results of previous works (\citealt{1993BaltA...2..407C, 2006ApJ...640..956M, 2017ApJS..232...23H, 2021ApJ...912..125G}). We found that the distribution of WMP shows a double-peak Gaussian-like structure, with peaks located around 249s \& 711s, respectively. This result is in agreement with the well-established correlation between the pulsation periods and amplitudes of \zcs~- short pulsation periods (100–300 s) typically have smaller amplitude, whereas long pulsation periods (600–1000 s) typically have larger amplitudes (\citealt{1982ApJ...258..651F,2006ApJ...640..956M}). We also tried to define any correlation between mass and WMP as predicted (\citealt{1982ApJ...258..651F}). Although the distribution is scattered, we can observe a trend in the extreme mass: lower-mass stars tend to have longer pulsation periods. The measurements between $\mobs$ and $\mses$ of \zcs~were generally consistent, as shown in Fig.\ref{fig:mass_compare}. The distribution of $\mses$ shows main peaks close to $\sim0.6M_\sun$ and a secondary peak $\sim0.85M_\sun$, which were already reported, e.g., the mass distribution of DA white dwarfs with $\teff \geq 7000$K (\citealt{2023MNRAS.518.5106J}). 

We defined a semi-empirical MRR with the trigonometrical $R_\pi$ and modeled mass. In general, the observational stellar parameters are consistent with the model. However, most of \zcs, with a mass close to $0.6M_\sun$, showed a larger $R_\pi$ than the predicted radius. This resulted in a discrepancy between the observation and the model, which is typically 5-10\%. However, for stars with $\mses>0.7M_\sun$, the discrepancy can reach $>15\%$, even considering the most extreme scenario for the ZZ Ceti - Blue edge and thick model (blue line in Fig. \ref{fig:MRR}). For the massive \zcs, the radius is almost fully dependent on the mass and not sensitive to the $\teff$ \citet{2019MNRAS.484.2711R}. Therefore, we attributed the large discrepancy to the potential difference between the $\mses$ and $\mobs$ for massive \zcs, which is in agreement with the parameters comparison between observational and seismological results.

The use of different models and instruments may introduce potential systematic uncertainties in the data and seismic analysis. However, instrumental bias can likely be neglected, as the pulsation frequencies of DAVs do not significantly change across different wavelength domains (\citealt{2006AAph...446..237D}). Instead, the main source of systematic differences stems from the models used in both seismic (\citealt{2018AJ....155..187B}) and spectroscopic (\citealt{2009A&A...498..517K}), and among different seismic models (\citealt{2006AAph...450..227C}). The systematic uncertainties from the spectroscopic analysis using different models are minor (e.g., refer to the recent work \citealt{2023arXiv230900239S}) compared to the discrepancies that arise from differences in the modeled atmospheres used in seismology and spectroscopy, as shown in Fig. \ref{fig:MRR}. These variations underscore the need for a more in-depth examination of the assumptions and limitations of each model to gain a deeper understanding of the underlying causes of the observed discrepancies.

Another potential source of systematic offset between the observed and theoretical MRR is the internal core composition assumed in evolutionary models.  Most calculations in the literature have adopted a "canonical" C/O mixture (typically 50/50 by mass), but variations in the central oxygen fraction can alter the white dwarf’s mean molecular weight and degeneracy pressure profile, hence its radius for a given mass.  For example, an oxygen-rich core (higher O/C) is more compact, leading to radii smaller by up to $\sim3\!-\!5\%$ compared to pure-carbon or balanced C/O models (e.g., \citealt{1997ApJ...486..413S,2001PASP..113..409F,2010A&ARv..18..471A}).  Conversely, a carbon-rich core produces a gentler MRR, with radii larger by a few percent.  
Moreover, for the most massive white dwarfs ($M>1.05\,M_\sun$), evolutionary models predict the formation of O/Ne cores rather than C/O.  O/Ne-core WDs have even higher mean molecular weights and thus are more compact—up to $\sim5\!-\!8\%$ smaller in radius at a given mass (e.g., \citealt{1996ApJ...460..489R,2015MNRAS.446.2599D}).  If some of our high-mass ZZ Ceti stars harbor O/Ne cores, then comparison against C/O-only grids would produce an apparent radius excess in the seismically determined MRR.  
If our sample harbors cores with systematically different C/O ratios (due to progenitor evolution, metallicity, or convective-boundary mixing), this could partially explain the $5\!-\!10\%$ radius excess at $\sim0.6\,M_\sun$ and the even larger deviations in the high-mass regime.  Future asteroseismic studies that directly constrain internal composition profiles, combined with updated evolutionary tracks spanning both C/O and O/Ne core mixtures, will be crucial to disentangle core-composition effects from other modeling uncertainties in the MRR.

Stars with infrared excess or being a component of a binary system are more likely to express larger luminosity, which could be a source of extra light for the observation. Although we had filtered out known targets with these two kinds, unsolved samples can exist, especially for those extreme outliers.  Another reason for the radius excess of massive \zcs~may come from the overestimated seismic mass under the assumption of Newtonian gravity (\citealt{2023MNRAS.524.5929C}). 
The seismic results in our collection utilized various models, and approximately half of the \ zcs~remained without analysis by asteroseismology. A comprehensive study with one updated model with no evolutionary model dependence (e.g., \citealt{2022FrASS...9.9045G}) would significantly benefit the constraint of the WD model.

%%Use section* for acknowledgements
\section*{Acknowledgements}
TC is supported by the LAMOST fellowship as a Youth Researcher, which is supported by the Special Funding for Advanced Users, budgeted and administrated by the Center for Astronomical Mega-Science, Chinese Academy of Sciences (CAMS), and acknowledges funding from the China Postdoctoral Science Foundation (2023M730297). TC and WKZ acknowledge funding from the support of the National Natural Science Foundation of China (NSFC) through grant 12273002. JNF acknowledges funding from the support of NSFC through grants 12090040/12090042/12427804, and the science research grants from the This work is supported by the China Manned Space Program with grant No. CMS-CSST-2025-A013, and the Central Guidance for Local Science and Technology Development Fund under No. ZYYD2025QY27. This paper made use of the Montreal White Dwarf Database \citealt{2017ASPC..509....3D} and the SIMBAD database, operated at CDS, Strasbourg, France.
\vspace{-1em}

%%use \balance somewhere in the left column of the last page to balance the two columns in the end page

%%References section
\bibliographystyle{apj}
\bibliography{references} 

%%Appendix
\clearpage
\appendix
\section{ZZ Cetis Table} \label{sec:data_tab}
Table \ref{tab:dav_info} includes stellar parameters of \zcs~in this work, including the GAIA DR3 ID and corresponding GAIA G magnitude and parallax. Effective temperature, $\logg$, and mass from the observation and seismological analysis can be found, respectively. The trigonometrical radius $R_\pi$ and the Mean weighted pulsating period (WMP) are also included. Uncertainties and collected oscillation properties can be found in the machine-readable tables.

\clearpage
\onecolumn
\begin{landscape} % Start Horizontal Page
\begin{longtable}{cccccccccccc}
\caption{Basic information for the \zcs in this work.\label{tab:dav_info}}\\
\hline
No. & GAIA ID & $\teff$  & $log~g$ & $M_{o}$  & $T_\mathrm{eff,ses}$ &  $M_{s}$ & G & Parallax & $R_{\pi}$ & WMP & Ref
\\
 &  & (K)  & (dex) & ($M_\sun$)  & (K) & ($M_\sun$)  & (mag) & (mas) & ($R_\sun/100$) & (s) & 
 \\
\hline
\endhead
\hline
\endfoot
\hline
\multicolumn{12}{p{1.2\textwidth}}{\textbf{References:} 1.\citet{2013ApJ...779...58R} 2.\citet{2007AAph...462..989C} 3.\citet{2022MNRAS.511.1574R} 4.\citet{2021ApJ...912..125G} 5.\citet{2005ApJ...625..966M}
6.\citet{2009MNRAS.396.1709C} 7.\citet{2019MNRAS.490.1803R} 8.\citet{1968ApJ...153..151L} 9.\citet{1972MNRAS.159P..25P} 10.\citet{1972MNRAS.156....1W} 11.\citet{1973BAAS....5...17F} 12.\citet{1998BaltA...7..153D}   13.\citet{2006AAph...446..237D} 
14.\citet{2012MNRAS.420.1462R} 15.\citet{2022RNAAS...6..107S} 16.\citet{2017MNRAS.468..239C} 17.\citet{2021AcA....71..281K} 18.\citet{2006AAph...450.1061V}
19.\citet{2021AAph...651A..14B} 20.\citet{2015ASPC..493..237G} 21.\citet{2020AJ....160..252V} 22.\citet{2006ApJ...640..956M} 23.\citet{2004ApJ...607..982M} 24.\citet{2006AAph...450..227C} 25.\citet{2006AJ....132..831G} 26.\citet{2017ApJS..232...23H} 27.\citet{2010MNRAS.405.2561C} 28.\citet{2016MNRAS.457.2855G} 29.\citet{2005AAph...442..629K} 30.\citet{2005AAph...443..195S} 31.\citet{2009MNRAS.399.1954B} 32.\citet{2013MNRAS.430...50C} 33.\citet{1976ApJ...209..853H} 34.\citet{1981ApJ...246..947V} 35.\citet{1981ApJ...246..947V} 36.\citet{1984ApJ...278..754K} 37.\citet{2004ApJ...600..404B} 38.\citet{1992ApJ...390L..89K} 39.\citet{1998BaltA...7..183K} 40.\citet{2000BaltA...9...87K} 41.\citet{2005AAph...432..219K} 42.\citet{2004ApJ...605L.133M} 43.\citet{2019AAph...632A.119C} 44.\citet{2013ApJ...771L...2H} 45.\citet{2012ApJ...750L..28H} 46.\citet{2019MNRAS.486.3560F} 47.\citet{2006MmSAI..77..486S} 48.\citet{2007ASPC..372..583V} 49.\citet{2019NewA...7301276C} 50.\citet{2017NewA...55...48L} 51.\citet{2001ApJ...557..792F} 52.\citet{2018MNRAS.478.2676B} 53.\citet{2014NewA...33...52S} 54.\citet{2014ApJ...789...85H} 55.\citet{2016MNRAS.455.2295G} 56.\citet{2003ApJ...591.1184F} 57.\citet{2002ApJ...580..429M} 58.\citet{2003ApJ...594..961M} 59.\citet{2013ApJ...771...17M} 60.\citet{2013ApJ...766...42H} 61.\citet{2011ApJ...730..128T} 62.\citet{2011ApJ...741L..16H} 63.\citet{1974ApJ...192..719R} 64.\citet{1976ApJ...205L.155M} 65.\citet{1982ApJ...254..676K} 66.\citet{1995BaltA...4..221K} 67.\citet{2008MNRAS.385..430C} 68.\citet{2012ApJ...757..177K} 69.\citet{1977ApJ...214L.123M} 70.\citet{1982MNRAS.200..563O} 71.\citet{2005ApJ...635.1239Y} 72.\citet{1983ApJ...271..744K} 73.\citet{1995ApJ...447..874K} 74.\citet{1976ApJ...207L..37R} 75.\citet{2016MNRAS.461.4059B} 76.\citet{2015ApJ...809...14B} 77.\citet{2017ApJ...847...34S} 78.\citet{1993MNRAS.263L..13S} 79.\citet{1995MNRAS.272L..21S} 80.\citet{1997fbs..conf..497S} 81.\citet{1971ApJ...163L..89L} 82.\citet{1977PASP...89..912S} 83.\citet{1980ApJ...240..865S} 84.\citet{1995BaltA...4..238K} 85.\citet{1976ApJ...210L..35M} 86.\citet{1986MNRAS.220P..19O} 87.\citet{1992MNRAS.258..415O} 88.\citet{1975ApJ...200L..89M} 89.\citet{2009JPhCS.172a2067T} 90.\citet{1981ApJ...250..349M} 91.\citet{1989AAph...215L..17V} 92.\citet{1998AAph...330..277J} 93.\citet{2001ASPC..226..313H} 94.\citet{2002MNRAS.335..399H} 95.\citet{2002AAph...388..219K} 96.\citet{2013MNRAS.429.1585F} 97.\citet{1983AAph...121L..23D} 98.\citet{1985ApJ...294..339F} 99.\citet{1997AAph...322..155V} 100.\citet{1999ASPC..169..129D} 101.\citet{2015MNRAS.449.3360L} 102.\citet{2007CoAst.150..251B} 103.\citet{2016ApJ...829...82B} 104.\citet{2014MNRAS.437.2566S} 105.\citet{1981AAph...103L..17V} 106.\citet{2002ASPC..259..378F} 107.\citet{1978ApJ...220..614R} 108.\citet{1984AJ.....89.1728F} 109.\citet{2008ApJ...676..573M} 110.\citet{2008MNRAS.388.1444H} 111.\citet{2012ApJ...751...91P} 112.\citet{2017ASPC..509..359P} 113.\citet{2014MNRAS.437.3183C} 114.\citet{1990ApJ...352L..45B} 115.\citet{1993AJ....106.1987B} 116.\citet{2015ApJ...815...56G} 117.\citet{2008CoAst.156...18H} 118.\citet{2000BaltA...9..133V} 119.\citet{1998BaltA...7..131G} 120.\citet{1995BaltA...4..245P} 121.\citet{1996AAph...314..182P} 122.\citet{2005ASPC..334..577H} 123.\citet{2013MNRAS.432..598P} 124.\citet{2014MNRAS.438.3086G} 125.\citet{2008PASP..120.1103S} 126.\citet{2013MNRAS.436.3573H} 127.\citet{2000ApJ...539..379K} 128.\citet{2004AAph...413..623C} 129.\citet{2004ApJ...610.1001T} 130.\citet{2012AJ....143..103S} 131.\citet{2005JRASC..99W.137G} 132.\citet{1987AAph...175L..13V} 133.\citet{1992AAph...264..547V} 134.\citet{2011MNRAS.415.1322P} 135.\citet{2014IAUS..301..469P} 136.\citet{1990ApJ...357..630W} 137.\citet{1991ApJ...374..330P} 138.\citet{1995PhDT........27K} 139.\citet{1998ApJ...495..424K} 140.\citet{2005ASPC..334..471T} 141.\citet{2015ASPC..493..217B} 
142.\citet{2020NatAs...4..690P} 
143.\citet{2022MNRAS.510..858R} 
144.\citet{2017NewA...55...48L}
145.\citet{2020AAph...638A..82B}
146.\citet{2023ApJ...948...74H}
147.\citet{2023MNRAS.518.1448R}
148.\citet{2023AAph...674A.204B}
149.\citet{2023MNRAS.523.1591G}
150.\citet{2023MNRAS.522.2181K}
151.\citet{2024MNRAS.528.5242G}
152.\citet{2023MNRAS.526.2846U}
153.\citet{2023MNRAS.524.5929C}
}
\endlastfoot
1 & 2449845669146456576 & 12140 & 8.19 & 0.72 & 11352 & 0.878 & 18.8277 & 4.3167 & 1.200 & 600.7 & 1,2 \\
2 & 5179395998203705344 & 10665 & 8.36 & 0.83 & 11176 & 0.917 & 18.7809 & 5.6866 & 1.106 & 900.88 & 1,2 \\
3 & 3074603225614954496 & 11832 & 8.12 & 0.67 & - & - & 18.2876 & 5.1835 & 1.252 & 264.35 & 2 \\
4 & 584401463934612352 & 10824 & 7.95 & 0.57 & 10992 & 0.800 & 17.3011 & 8.3665 & 1.368 & 786.76 & 1,2,3 \\
5 & 812202337427426304 & 11667 & 8.05 & 0.64 & - & - & 16.6199 & 10.9835 & 1.315 & 475.41 & 2,4 \\
6 & 1311359652701512192 & 10757 & 8.41 & 0.86 & 11168 & 1.024 & 18.1507 & 7.7440 & 1.073 & 339.06 & 1,2 \\
7 & 2646176183389736192 & 11400 & 7.99 & 0.60 & - & - & 19.3675 & 3.4960 & 1.264 & 923.15 & 2 \\
8 & 2545567334691271424 & 11805 & 8.00 & 0.60 & 11600 & 0.630 & 17.4402 & 7.1882 & 1.389 & 216.33 & 5,6 \\
9 & 2781124708664855040 & 11758 & 8.36 & 0.83 & 11472 & 0.949 & 18.7110 & 5.2728 & 1.134 & 571.03 & 1,5 \\
10 & 673438365068121600 & 11777 & 7.99 & 0.60 & 11104 & 0.670 & 18.3338 & 4.7774 & 1.399 & 199.56 & 5,6 \\
11 & 900922102470856576 & 12285 & 8.04 & 0.63 & 11400 & 0.650 & 17.4465 & 7.1677 & 1.317 & 226.15 & 5,6 \\
12 & 815420470522957568 & 11474 & 7.91 & 0.56 & 11696 & 0.530 & 17.7638 & 6.0893 & 1.431 & 281.88 & 5,6 \\
13 & 1046810046887551744 & 12213 & 7.84 & 0.52 & 12304 & 0.530 & 18.3547 & 4.1815 & 1.535 & 284.14 & 5 \\
14 & 851543000907468672 & 10956 & 7.85 & 0.52 & 11600 & 0.530 & 18.9084 & 3.6285 & 1.549 & 267.25 & 5 \\
15 & 837550886514384128 & 11290 & 7.91 & 0.55 & - & - & 17.9696 & 5.5458 & 1.513 & 741.91 & 5,6 \\
16 & 1561336305631985152 & 11116 & 7.95 & 0.58 & - & - & 18.6503 & 4.3004 & 1.474 & 324.0 & 5,6 \\
17 & 1767911240292360704 & 10853 & 8.37 & 0.84 & 11672 & 0.976 & 18.9592 & 5.0997 & 1.073 & 739.8 & 1,5,7 \\
18 & 2678581436759837184 & 11535 & 8.17 & 0.71 & 11632 & 0.878 & 17.8951 & 6.7164 & 1.232 & 245.61 & 1,5,7 \\
19 & 152371871863155840 & 11363 & 7.86 & 0.52 & 11112 & 0.548 & 15.0304 & 20.7244 & 1.451 & 655.94 & 4,8,9,10,11,12,13,14 \\
20 & 1218748892101806720 & 10853 & 8.29 & 0.78 & - & - & 17.6021 & 8.9112 & 1.168 & 572.2 & 15,16 \\
21 & 4460327252045435648 & 11887 & 8.04 & 0.63 & 11488 & 0.609 & 16.2465 & 12.8495 & 1.279 & 663.17 & 14,17,18 \\
22 & 1900545847646195840 & 9258 & 7.24 & 0.27 & 11400 & 0.460 & 16.1965 & 10.9800 & 2.110 & 1141.0 & 19 \\
23 & 267146523434846208 & 10967 & 8.02 & 0.62 & 11904 & 0.700 & 15.2442 & 21.7164 & 1.300 & 852.1 & 19,20 \\
24 & 2323704339384503040 & 10716 & 8.00 & 0.60 & - & - & 15.8789 & 16.3232 & 1.344 & 1152.4 & 25 \\
25 & 599947802437634176 & 11754 & 7.99 & 0.60 & 11504 & 0.646 & 17.6794 & 6.4855 & 1.363 & 248.15 & 7 \\
26 & 3879246725541954816 & 11432 & 7.95 & 0.57 & 11456 & 0.705 & 17.5515 & 6.7979 & 1.436 & 200.93 & 7 \\
27 & 1199705488147772544 & 11354 & 7.76 & 0.47 & 12208 & 0.705 & 16.6323 & 9.5122 & 1.536 & 536.49 & 7 \\
28 & 6681773947733560192 & 11061 & 8.03 & 0.62 & 10800 & 0.593 & 13.7459 & 42.8354 & 1.297 & 787.87 & 4,7 \\
29 & 4933533142262051968 & 11097 & 8.08 & 0.65 & 11576 & 0.632 & 15.4586 & 20.0610 & 1.271 & 762.85 & 1,6,7,14,33 \\
30 & 3852118307646195328 & 10762 & 8.18 & 0.71 & 11384 & 0.675 & 15.2510 & 24.5566 & 1.169 & 1236.47 & 1,7,27 \\
31 & 3705386281897262848 & 11772 & 8.08 & 0.65 & 11520 & 0.609 & 16.0315 & 14.5934 & 1.295 & 290.9 & 7,14,18 \\
32 & 3662011643397066112 & 11609 & 8.00 & 0.60 & 11680 & 0.686 & 16.7596 & 9.9260 & 1.387 & 204.21 & 7,23 \\
33 & 3642258577702062720 & 10827 & 7.91 & 0.55 & 11408 & 0.548 & 16.0243 & 14.2810 & 1.490 & 850.2 & 7,14,30 \\
34 & 4451438628247615744 & 12108 & 8.39 & 0.85 & 11816 & 0.705 & 17.8059 & 7.7736 & 1.033 & 115.61 & 1,7,28 \\
35 & 2026653131220381824 & 11844 & 8.02 & 0.61 & 11568 & 0.632 & 15.1274 & 21.0917 & 1.292 & 280.45 & 6,7,14,34,35,36,37 \\
36 & 2641630119421561856 & 9625 & 7.36 & 0.31 & 11080 & 1.024 & 18.1003 & 4.6053 & 2.104 & 316.3 & 1,7,22,23 \\
37 & 6127333286605955072 & 11003 & 8.65 & 1.01 & 11648 & 1.160 & 13.7929 & 67.4058 & 0.812 & 584.76 & 38,39,40,41,42,43 \\
38 & 1635687790163070976 & 11422 & 8.83 & 1.11 & 12064 & 1.220 & 17.2439 & 15.4809 & 0.711 & 471.51 & 43,44 \\
39 & 1027876005685373696 & 11947 & 8.59 & 0.98 & 12048 & 1.140 & 18.3151 & 7.3877 & 0.865 & 444.76 & 16,43 \\
40 & 3810933247769901696 & 12206 & 8.04 & 0.63 & 12400 & 0.630 & 14.6330 & 26.1508 & 1.280 & 134.03 & 46,47 \\
41 & 361418474306342656 & 11629 & 7.98 & 0.59 & 12064 & 0.570 & 16.3901 & 11.6709 & 1.342 & 212.7 & 14,25,49 \\
42 & 136610338318092544 & 10981 & 7.99 & 0.60 & 11696 & 0.980 & 16.0612 & 14.4783 & 1.341 & 958.33 & 14,31,50,51 \\
43 & 725488146015775744 & 11409 & 8.42 & 0.87 & - & - & 18.7106 & 5.7277 & 0.978 & 498.5 & 16 \\
44 & 2290663907599394560 & 12419 & 8.21 & 0.74 & - & - & 19.0956 & 3.7710 & 1.261 & 203.7 & 16 \\
45 & 1566603962760532736 & 10300 & 7.68 & 0.43 & 11904 & 0.800 & 14.0682 & 32.4824 & 1.640 & 783.92 & 52 \\
46 & 1326398777041821568 & 9900 & 8.00 & 0.60 & - & - & 14.6169 & 32.7870 & 1.314 & 751.0 & 52 \\
47 & 4921390960477978112 & 11610 & 7.90 & 0.55 & - & - & 15.7742 & 14.7234 & 1.435 & 291.5 & 24 \\
48 & 2497677379892554240 & 9918 & 7.80 & 0.49 & 11176 & 0.632 & 18.8562 & 4.1521 & 1.628 & 1029.0 & 1,24 \\
49 & 3690568537350703744 & 12446 & 8.18 & 0.72 & 11168 & 0.705 & 18.6651 & 4.3782 & 1.280 & 785.32 & 1,24 \\
50 & 4415799779897925760 & 11208 & 7.91 & 0.55 & 11128 & 0.632 & 16.5130 & 10.9336 & 1.450 & 276.24 & 3,24 \\
51 & 4407307731905788032 & 10864 & 8.16 & 0.69 & - & - & 19.4249 & 3.1183 & 1.524 & 644.0 & 24 \\
52 & 1326864347200777216 & 11067 & 8.01 & 0.61 & 11496 & 0.721 & 19.0917 & 3.6654 & 1.446 & 809.3 & 1,24 \\
53 & 2687661650459472000 & 11660 & 8.12 & 0.68 & 11568 & 0.976 & 18.0092 & 6.0839 & 1.238 & 290.87 & 1,24 \\
54 & 6894920835191201536 & 11255 & 7.88 & 0.54 & - & - & 18.6952 & 3.9682 & 1.471 & 445.18 & 24 \\
55 & 2666848037078840704 & 13876 & 8.03 & 0.63 & - & - & 18.5061 & 4.0645 & 1.312 & 210.2 & 24 \\
56 & 2731334870789219456 & 11634 & 7.98 & 0.59 & - & - & 18.6100 & 4.2855 & 1.362 & 589.41 & 24 \\
57 & 2607171317631314432 & 11163 & 7.97 & 0.59 & - & - & 18.8057 & 4.0521 & 1.407 & 1016.92 & 24 \\
58 & 3277341526122556160 & 10048 & 7.10 & 0.24 & - & - & 16.5977 & 7.1846 & 2.465 & 1340.55 & 25,53 \\
59 & 2439184705619919488 & 10602 & 7.98 & 0.59 & 11184 & 0.737 & 13.2931 & 53.5809 & 1.343 & 1074.72 & 25,26,54,146 \\
60 & 6617454235493798272 & 11544 & 7.95 & 0.58 & 11848 & 0.632 & 16.0699 & 13.3920 & 1.372 & 260.8 & 14,25 \\
61 & 3571559292842744960 & 11681 & 8.02 & 0.62 & 12176 & 0.705 & 16.0306 & 14.0332 & 1.396 & 235.05 & 14,25,48 \\
62 & 2534336854204440320 & 11517 & 7.79 & 0.49 & 11696 & 0.710 & 16.9349 & 8.0745 & 1.612 & 179.95 & 26,27,146 \\
63 & 2458857988002637952 & 11666 & 7.92 & 0.56 & 11296 & 0.610 & 18.4160 & 4.4493 & 1.525 & 173.28 & 27 \\
64 & 3562411351802430592 & 12079 & 8.07 & 0.65 & 11336 & 0.837 & 17.6008 & 6.8893 & 1.337 & 231.69 & 1,27 \\
65 & 3662700556150949376 & 11768 & 7.97 & 0.59 & 11400 & 0.695 & 17.1519 & 8.0202 & 1.440 & 172.94 & 27 \\
66 & 2681074029619738624 & 11992 & 7.92 & 0.57 & 11800 & 0.585 & 19.0216 & 3.2210 & 1.471 & 199.8 & 27 \\
67 & 2549166547349010560 & 11809 & 8.00 & 0.61 & 12096 & 0.585 & 18.7462 & 3.9947 & 1.436 & 258.2 & 27 \\
68 & 3793823193278661888 & 12348 & 8.00 & 0.61 & 11800 & 0.580 & 17.8528 & 5.7161 & 1.410 & 234.64 & 26,27,146 \\
69 & 3836338421707173504 & 12126 & 8.05 & 0.63 & 11400 & 0.640 & 18.2106 & 5.1253 & 1.354 & 251.78 & 27 \\
70 & 2615874708080122240 & 10970 & 8.18 & 0.72 & 11944 & 0.949 & 18.5872 & 5.2348 & 1.213 & 777.02 & 1,2,27 \\
71 & 6918986155511040 & 11319 & 7.96 & 0.58 & - & - & 18.1124 & 5.4946 & 1.488 & 300.83 & 27 \\
72 & 2536148746287884160 & 10483 & 7.83 & 0.51 & 10984 & 0.660 & 18.1966 & 5.2646 & 1.561 & 860.73 & 1,22,23,55 \\
73 & 2534803111558973056 & 12099 & 7.94 & 0.57 & 11824 & 0.800 & 18.8283 & 3.6937 & 1.437 & 277.81 & 1,6,23,26,60,61,146 \\
74 & 36207059639735296 & 11951 & 8.25 & 0.76 & 11728 & 0.878 & 16.7103 & 11.7143 & 1.168 & 184.5 & 1,32 \\
75 & 4766810380210581632 & 11251 & 8.40 & 0.86 & 11280 & 0.949 & 15.9318 & 20.1009 & 1.019 & 686.92 & 1,6,56 \\
76 & 3092047767863216256 & 11560 & 8.10 & 0.67 & 11416 & 0.770 & 17.5521 & 7.4888 & 1.246 & 619.52 & 1,29 \\
77 & 914776838038486784 & 11379 & 8.39 & 0.85 & 11920 & 0.998 & 18.5660 & 5.8746 & 1.039 & 636.62 & 1,23 \\
78 & 582222433752256384 & 10941 & 8.03 & 0.62 & 10992 & 0.837 & 17.8354 & 6.5918 & 1.341 & 940.27 & 1,29 \\
79 & 3844065450124684544 & 11197 & 8.57 & 0.97 & 11120 & 0.949 & 18.4017 & 7.4170 & 0.886 & 729.0 & 1,22,23 \\
80 & 798566915774818176 & 11599 & 7.99 & 0.60 & 11984 & 0.593 & 15.5395 & 17.3947 & 1.371 & 509.1 & 1,14,37,63,64,65,66,67 \\
81 & 1025139218164253568 & 11822 & 8.10 & 0.67 & 11672 & 0.770 & 18.7537 & 4.2039 & 1.325 & 159.19 & 1,23 \\
82 & 3840420221186254208 & 10347 & 7.58 & 0.39 & 10816 & 0.770 & 18.1897 & 4.7018 & 1.731 & 255.25 & 1,32 \\
83 & 3601117219117585920 & 11198 & 8.02 & 0.61 & 11712 & 0.917 & 18.2066 & 5.3675 & 1.364 & 288.0 & 1,32 \\
84 & 3905315497697270400 & 11344 & 8.30 & 0.78 & 11656 & 0.770 & 21.7595 & - & - & 718.25 & 1,6,29 \\
85 & 3698504571761639808 & 10676 & 7.73 & 0.46 & 11320 & 0.705 & 18.5283 & 4.1532 & 1.689 & 221.08 & 1,29 \\
86 & 3693488019896536704 & 11210 & 8.06 & 0.64 & 11184 & 0.837 & 16.7886 & 10.6876 & 1.308 & 443.29 & 1,29 \\
87 & 3687319308692216832 & 11674 & 8.33 & 0.81 & 11536 & 0.917 & 18.5491 & 5.5267 & 1.143 & 632.73 & 1,68 \\
88 & 3663471382521565184 & 11316 & 8.46 & 0.90 & 11248 & 0.837 & 18.6013 & 6.1358 & 1.014 & 715.0 & 1,29 \\
89 & 5772718006135360128 & 11819 & 8.08 & 0.66 & 12032 & 0.705 & 13.4242 & 47.8778 & 1.250 & 169.89 & 1,14,69,70,71 \\
90 & 1431783457574556672 & 11848 & 8.25 & 0.76 & 11280 & 0.976 & 12.2790 & 91.3400 & 1.100 & 109.29 & 1,14,29,72,73 \\
91 & 1632472715084508800 & 10975 & 8.38 & 0.84 & 11272 & 1.023 & 16.9473 & 12.6324 & 1.046 & 752.1 & 1,6,22,23 \\
92 & 2092134443120038528 & 11769 & 8.33 & 0.81 & 12032 & 0.837 & 14.6341 & 32.9424 & 1.042 & 411.67 & 1,6,14,74,75 \\
93 & 2101031858707484288 & 10621 & 8.09 & 0.66 & 11392 & 0.837 & 17.7037 & 7.5671 & 1.260 & 861.25 & 1,62,76,26146 \\
94 & 2720714898429962624 & 11602 & 8.40 & 0.85 & 11136 & 0.949 & 17.9720 & 7.6653 & 1.016 & 714.8 & 1,32 \\
95 & 1992360153807695104 & 10222 & 8.21 & 0.73 & 11752 & 0.976 & 19.2932 & 4.3078 & 1.115 & 601.2 & 1 \\
96 & 2796821959433536640 & 11393 & 7.95 & 0.57 & - & - & 17.7653 & 6.2307 & 1.434 & 855.95 & 4 \\
97 & 89586833041913216 & 9774 & 7.81 & 0.49 & - & - & 18.3290 & 5.4713 & 1.556 & 811.48 & 4 \\
98 & 516606710842437248 & 11097 & 7.98 & 0.59 & - & - & 15.5835 & 17.6745 & 1.352 & 667.72 & 3,4 \\
99 & 462506821746606464 & 10960 & 8.03 & 0.62 & - & - & 16.0729 & 14.9050 & 1.303 & 849.98 & 4,21 \\
100 & 258439731372229120 & 11110 & 7.95 & 0.57 & 11168 & 0.686 & 15.8700 & 15.3532 & 1.365 & 829.42 & 3,4,21 \\
101 & 3181589319065856384 & 11716 & 8.51 & 0.93 & 12704 & 0.820 & 16.2602 & 18.1966 & 0.901 & 508.86 & 3,4,21 \\
102 & 3224908977688888064 & 10936 & 8.00 & 0.60 & - & - & 16.1260 & 14.2880 & 1.335 & 873.98 & 4,21 \\
103 & 996792884983596800 & 10628 & 8.03 & 0.62 & - & - & 16.6608 & 11.7989 & 1.307 & 861.5 & 4 \\
104 & 3117810051152885760 & 10304 & 7.79 & 0.49 & - & - & 17.3373 & 7.9178 & 1.507 & 765.3 & 4 \\
105 & 750807768499445376 & 11469 & 8.16 & 0.70 & - & - & 17.5621 & 7.8409 & 1.250 & 629.09 & 4 \\
106 & 3983606596814071680 & 12877 & 8.86 & 1.13 & - & - & 17.5180 & 12.3301 & 0.705 & 420.15 & 4 \\
107 & 1485679322841561728 & 10821 & 7.90 & 0.55 & - & - & 17.0630 & 8.7667 & 1.502 & 894.07 & 4 \\
108 & 1325292217371985024 & 10995 & 7.91 & 0.55 & 10968 & 0.675 & 16.7242 & 10.2380 & 1.470 & 771.27 & 3,4 \\
109 & 4128367507456250240 & 11349 & 8.08 & 0.65 & - & - & 16.2949 & 13.3472 & 1.259 & 529.45 & 4 \\
110 & 4570546317703725312 & 11361 & 8.04 & 0.63 & 11640 & 0.750 & 16.0906 & 14.2690 & 1.305 & 569.25 & 4,21,151 \\
111 & 4611476737557895552 & 11010 & 7.87 & 0.53 & - & - & 16.9489 & 8.8667 & 1.543 & 740.5 & 4 \\
112 & 2114811453822316160 & 12442 & 8.50 & 0.92 & 11712 & 0.660 & 16.2436 & 17.2120 & 0.919 & 437.77 & 3,4,21 \\
113 & 4586334819344859392 & 10797 & 8.00 & 0.60 & - & - & 17.9364 & 6.3001 & 1.372 & 610.43 & 4 \\
114 & 4517992170175866368 & 10987 & 7.95 & 0.57 & - & - & 16.7150 & 10.4368 & 1.389 & 496.69 & 4 \\
115 & 2143272965142309888 & 10556 & 7.97 & 0.58 & - & - & 17.3345 & 8.3410 & 1.432 & 1184.59 & 4 \\
116 & 4217793816094052480 & 11551 & 8.44 & 0.88 & - & - & 15.2050 & 28.0908 & 0.965 & 707.17 & 4 \\
117 & 1862796379351925888 & 10050 & 7.37 & 0.31 & - & - & 17.4627 & 5.7324 & 2.005 & 1325.49 & 4 \\
118 & 4226824448690266496 & 11208 & 7.94 & 0.57 & - & - & 18.1965 & 5.1277 & 1.404 & 785.68 & 4 \\
119 & 1974423541446133248 & 11539 & 8.12 & 0.68 & - & - & 17.7741 & 6.8134 & 1.238 & 831.17 & 4 \\
120 & 1893512958955407232 & 9739 & 7.36 & 0.31 & - & - & 17.7449 & 5.2690 & 2.020 & 733.58 & 4 \\
121 & 2769930207121379968 & 10537 & 7.88 & 0.54 & 10952 & 0.548 & 16.0623 & 14.1806 & 1.481 & 1061.78 & 4,14,57 \\
122 & 2850846184091080576 & 11174 & 8.01 & 0.61 & - & - & 17.5541 & 7.2332 & 1.414 & 865.67 & 4 \\
123 & 5259226009874614912 & 10997 & 8.00 & 0.60 & - & - & 16.8016 & 10.3414 & 1.342 & 746.31 & 4 \\
124 & 5349284155459347968 & 10784 & 7.95 & 0.57 & - & - & 16.4008 & 12.3840 & 1.363 & 788.31 & 4 \\
125 & 5341381621826727680 & 11288 & 8.03 & 0.62 & - & - & 16.8426 & 10.1467 & 1.287 & 890.5 & 4 \\
126 & 5961193055261256320 & 11132 & 8.04 & 0.63 & 11376 & 0.660 & 13.5562 & 46.8046 & 1.284 & 608.54 & 3,4 \\
127 & 6464503543777416576 & 11098 & 7.95 & 0.58 & 11392 & 0.686 & 15.9196 & 15.0544 & 1.360 & 862.19 & 3,4 \\
128 & 2611423167751076224 & 11255 & 8.07 & 0.65 & - & - & 16.9197 & 9.9904 & 1.322 & 529.74 & 4 \\
129 & 360447124498490240 & 10344 & 7.50 & 0.36 & - & - & 17.5188 & 5.8323 & 1.961 & 800.45 & 4 \\
130 & 2539057401220177920 & 11068 & 7.89 & 0.54 & 11480 & 0.679 & 16.7093 & 9.9405 & 1.482 & 541.64 & 4,146 \\
131 & 3356726158272880768 & 11332 & 8.06 & 0.64 & - & - & 17.4643 & 7.7354 & 1.270 & 796.18 & 4 \\
132 & 986383460511639040 & 11400 & 7.98 & 0.59 & - & - & 17.4618 & 7.2791 & 1.408 & 627.4 & 4 \\
133 & 1617418957790376576 & 11822 & 8.05 & 0.63 & - & - & 17.1781 & 8.3970 & 1.363 & 458.12 & 3,4 \\
134 & 1601328670269782400 & 10126 & 7.38 & 0.32 & 10968 & 0.548 & 17.0913 & 6.8065 & 2.123 & 721.53 & 3,4,20 \\
135 & 1338943341425719168 & 11588 & 7.96 & 0.58 & - & - & 17.3374 & 7.4281 & 1.508 & 779.78 & 4,22,23 \\
136 & 1944856063168152832 & 11388 & 7.99 & 0.60 & - & - & 16.8045 & 9.9252 & 1.327 & 1050.6 & 4 \\
137 & 2823262637102613248 & 11271 & 7.96 & 0.58 & - & - & 17.4059 & 7.4091 & 1.440 & 364.2 & 4 \\
138 & 5435944535217680640 & 11258 & 7.55 & 0.38 & 11192 & 0.493 & 15.9774 & 11.1925 & 1.753 & 464.8 & 3 \\
139 & 5757034091238475520 & 13914 & 8.21 & 0.74 & 12768 & 0.705 & 16.4651 & 11.6247 & 1.114 & 415.21 & 3 \\
140 & 2120456419536355456 & 12439 & 7.23 & 0.28 & - & - & 16.2964 & 7.3505 & 2.226 & 1074.51 & 3 \\
141 & 3015008531455357824 & 11798 & 7.84 & 0.52 & 12496 & 0.542 & 14.7341 & 22.8983 & 1.436 & 262.65 & 3 \\
142 & 2450585266810809216 & 10895 & 8.01 & 0.61 & 11368 & 0.609 & 15.2513 & 21.6795 & 1.331 & 743.18 & 3 \\
143 & 4627855367706529152 & 10792 & 7.63 & 0.41 & 12528 & 0.609 & 16.5225 & 9.6891 & 1.823 & 413.87 & 3 \\
144 & 5741503313402334976 & 12012 & 8.00 & 0.60 & 11928 & 0.686 & 16.7730 & 9.8843 & 1.286 & 415.81 & 3 \\
145 & 4665237908354588288 & 11706 & 7.94 & 0.57 & 12600 & 0.570 & 14.9907 & 21.6278 & 1.374 & 236.51 & 3 \\
146 & 6602158345124764032 & 11557 & 7.97 & 0.59 & 11608 & 0.542 & 15.6098 & 17.0969 & 1.325 & 336.66 & 3 \\
147 & 5617205933369326592 & 13079 & 7.41 & 0.34 & 11480 & 0.542 & 14.4457 & 18.9025 & 1.840 & 278.17 & 3 \\
148 & 5020319141229055360 & 10380 & 7.16 & 0.26 & - & - & 15.9206 & 9.9045 & 2.342 & 893.97 & 3 \\
149 & 436085007572835072 & 11374 & 8.00 & 0.60 & 11592 & 0.609 & 16.3260 & 12.4541 & 1.326 & 369.4 & 3,21 \\
150 & 4974784825671467648 & 10207 & 7.23 & 0.27 & - & - & 16.5958 & 7.7280 & 2.207 & 1457.92 & 3 \\
151 & 5566825554660211200 & 13936 & 7.48 & 0.36 & - & - & 16.0972 & 8.4924 & 1.961 & 268.45 & 3 \\
152 & 5470271185153118208 & 9868 & 7.52 & 0.36 & 10640 & 0.493 & 16.8470 & 8.7053 & 1.916 & 644.75 & 3 \\
153 & 4733604373137759616 & 11100 & 8.02 & 0.61 & 10952 & 0.570 & 14.6223 & 28.6836 & 1.294 & 900.84 & 3 \\
154 & 4721093923677909376 & 12345 & 8.26 & 0.76 & 11688 & 0.686 & 15.8772 & 16.6841 & 1.136 & 307.21 & 3 \\
155 & 2260805780286092032 & 16317 & 7.52 & 0.39 & 11312 & 0.493 & 16.0486 & 7.7236 & 1.904 & 741.36 & 3 \\
156 & 2240031951187372928 & 11324 & 7.97 & 0.59 & 11912 & 0.593 & 16.4472 & 11.7266 & 1.383 & 214.59 & 3,21 \\
157 & 2155960371551164416 & 11148 & 8.06 & 0.64 & 11408 & 0.525 & 15.0372 & 24.2794 & 1.270 & 739.22 & 3,21 \\
158 & 4741117370449721216 & 11466 & 8.06 & 0.64 & 11296 & 0.570 & 16.4573 & 12.2320 & 1.301 & 677.27 & 3 \\
159 & 1669095729418204416 & 15658 & 8.01 & 0.62 & 12016 & 0.660 & 16.1447 & 10.7088 & 1.367 & 282.66 & 3 \\
160 & 6402648970968695424 & 11464 & 7.93 & 0.57 & 11632 & 0.690 & 16.1174 & 13.0456 & 1.399 & 270.26 & 3 \\
161 & 3937174946624964224 & 11851 & 8.02 & 0.62 & 11712 & 0.493 & 16.2995 & 12.3179 & 1.331 & 271.12 & 3,21 \\
162 & 1897597369775277568 & 11355 & 7.91 & 0.55 & 11320 & 0.570 & 15.9617 & 14.0391 & 1.395 & 325.61 & 3,21 \\
163 & 5246823346920176128 & 11416 & 7.97 & 0.59 & 12016 & 0.542 & 16.1183 & 13.3881 & 1.369 & 687.19 & 3 \\
164 & 1712016196599965312 & 9056 & 7.49 & 0.38 & - & - & 15.5890 & 17.1382 & 1.868 & 386.08 & 3 \\
165 & 6350177973988575232 & 11494 & 8.00 & 0.61 & 11904 & 0.621 & 16.4937 & 11.4285 & 1.421 & 702.0 & 3 \\
166 & 5802990619265616640 & 11482 & 7.96 & 0.58 & 11312 & 0.548 & 15.8691 & 14.9444 & 1.349 & 287.85 & 3 \\
167 & 6365271657299575680 & 11073 & 7.13 & 0.25 & - & - & 15.4161 & 11.4201 & 2.305 & 1036.98 & 3 \\
168 & 6461421956281723392 & 11192 & 8.23 & 0.75 & 11008 & 0.770 & 14.6479 & 18.4533 & 1.974 & 690.65 & 3 \\
169 & 2055661546498684416 & 11444 & 8.38 & 0.85 & 11552 & 0.745 & 15.6775 & 22.0790 & 0.994 & 549.7 & 3,21 \\
170 & 6410501923531836928 & 11464 & 7.99 & 0.60 & 11568 & 0.593 & 15.8112 & 15.6906 & 1.327 & 309.81 & 3 \\
171 & 3301217592917972864 & 11558 & 8.04 & 0.63 & 10096 & 0.609 & 16.5579 & 11.7172 & 1.286 & 820.81 & 3 \\
172 & 3494580937593244672 & 11386 & 8.02 & 0.62 & 11504 & 0.609 & 16.0329 & 14.6097 & 1.383 & 298.14 & 3,4 \\
173 & 5152748715429796096 & 11150 & 8.02 & 0.62 & 11264 & 0.705 & 16.6826 & 10.9675 & 1.376 & 808.65 & 3 \\
174 & 2970066126811876992 & 11869 & 8.42 & 0.87 & 12120 & 0.837 & 16.5335 & 14.8600 & 0.964 & 545.48 & 3 \\
175 & 2371650330620404224 & 10722 & 7.88 & 0.54 & 10968 & 0.609 & 17.3830 & 7.6126 & 1.535 & 812.23 & 3 \\
176 & 4687985493966045696 & 11278 & 7.93 & 0.56 & 11400 & 0.609 & 16.9722 & 8.9625 & 1.433 & 566.2 & 3 \\
177 & 4699906742631457280 & 11689 & 8.02 & 0.61 & 11552 & 0.550 & 17.0044 & 9.1786 & 1.339 & 363.14 & 3 \\
178 & 161359000733439488 & 10426 & 7.88 & 0.54 & 11600 & 0.570 & 17.3960 & 8.0542 & 1.542 & 626.41 & 3 \\
179 & 4783648885393612160 & 10816 & 7.90 & 0.55 & 10896 & 0.609 & 17.0540 & 8.9552 & 1.497 & 800.25 & 3 \\
180 & 4869262701888061056 & 11678 & 8.05 & 0.63 & 11128 & 0.639 & 17.1534 & 8.6702 & 1.376 & 837.44 & 3 \\
181 & 3126624316184360576 & 10956 & 7.99 & 0.60 & 10816 & 0.646 & 16.2260 & 13.9320 & 1.292 & 618.51 & 3 \\
182 & 4621529705513856512 & 9748 & 7.57 & 0.38 & 10912 & 0.542 & 17.4995 & 6.8309 & 1.961 & 290.18 & 3 \\
183 & 4660769493122896128 & 12007 & 8.01 & 0.61 & 12392 & 0.660 & 16.9051 & 9.0375 & 1.387 & 337.95 & 3 \\
184 & 5215511523500404352 & 15499 & 8.15 & 0.68 & 10992 & 0.609 & 16.9351 & 7.9146 & 1.345 & 1007.21 & 3 \\
185 & 5216388693260726016 & 11239 & 7.96 & 0.58 & 11776 & 0.609 & 17.4274 & 7.5123 & 1.485 & 447.23 & 3 \\
186 & 3462095007557034368 & 11092 & 7.97 & 0.58 & 10848 & 0.660 & 17.0583 & 9.0329 & 1.367 & 746.53 & 3 \\
187 & 1699066870202540288 & 11243 & 7.65 & 0.42 & 11376 & 0.542 & 17.0147 & 7.4564 & 1.755 & 709.34 & 3 \\
188 & 5820030266373269632 & 10809 & 7.86 & 0.53 & 11312 & 0.579 & 17.0193 & 8.6491 & 1.464 & 1227.25 & 3 \\
189 & 6105527256602819200 & 11006 & 7.99 & 0.60 & 10944 & 0.690 & 17.4130 & 7.8885 & 1.321 & 696.27 & 3 \\
190 & 1624129483413023360 & 11081 & 7.96 & 0.58 & 11472 & 0.609 & 17.1418 & 8.5738 & 1.448 & 840.92 & 3 \\
191 & 4566554472019360128 & 10700 & 8.03 & 0.62 & 10712 & 0.609 & 17.0114 & 10.0378 & 1.313 & 769.06 & 3 \\
192 & 6468707079810027392 & 10876 & 7.95 & 0.57 & 11112 & 0.609 & 16.6454 & 11.0062 & 1.370 & 909.04 & 3 \\
193 & 6482779145017901184 & 11151 & 7.96 & 0.58 & 11248 & 0.609 & 17.3960 & 7.6831 & 1.353 & 557.26 & 3 \\
194 & 6403387804126326144 & 11272 & 7.98 & 0.59 & 11240 & 0.525 & 17.3011 & 7.9749 & 1.372 & 696.3 & 3 \\
195 & 6519720296173216512 & 11048 & 7.92 & 0.56 & 11032 & 0.705 & 16.8602 & 9.4819 & 1.451 & 865.74 & 3 \\
196 & 362222801417874816 & 10537 & 7.09 & 0.24 & - & - & 16.3139 & 7.6965 & 2.371 & 473.0 & 77 \\
197 & 3431829747413234688 & 11002 & 8.00 & 0.60 & - & - & 17.6093 & 7.2084 & 1.322 & 830.0 & 77 \\
198 & 376478863024306688 & 10560 & 7.85 & 0.52 & - & - & 18.3402 & 4.8788 & 1.497 & 1174.0 & 77 \\
199 & 296688816025094144 & 11148 & 7.97 & 0.59 & - & - & 18.5830 & 4.4993 & 1.501 & 310.0 & 77 \\
200 & 2863526233218817024 & 10308 & 7.89 & 0.54 & - & - & 16.7020 & 11.1360 & 1.410 & 1459.0 & 21 \\
201 & 2779284538516313600 & 10884 & 8.00 & 0.60 & - & - & 16.4267 & 12.6120 & 1.383 & 1579.0 & 21 \\
202 & 2789405753503977472 & 10520 & 8.01 & 0.60 & - & - & 16.9660 & 10.1400 & 1.362 & 1102.0 & 21 \\
203 & 302143768088623488 & 11677 & 8.14 & 0.69 & - & - & 16.3730 & 13.1750 & 1.202 & 143.0 & 21 \\
204 & 575585919005741184 & 11288 & 8.73 & 1.06 & - & - & 17.8200 & 11.0530 & 0.779 & 330.0 & 21 \\
205 & 3400048535611299456 & 11686 & 7.94 & 0.57 & - & - & 16.4330 & 11.2300 & 1.364 & 196.0 & 21 \\
206 & 192275966334956672 & 13501 & 8.91 & 1.16 & - & - & 16.3670 & 21.5310 & 0.630 & 809.0 & 21 \\
207 & 3458597083113101952 & 11674 & 7.88 & 0.54 & - & - & 16.4320 & 10.8210 & 1.418 & 256.0 & 21 \\
208 & 3169486960220617088 & 11555 & 8.30 & 0.80 & - & - & 15.0831 & 26.9490 & 1.055 & 491.0 & 21 \\
209 & 983538336734107392 & 12376 & 8.03 & 0.62 & - & - & 16.6750 & 10.1900 & 1.323 & 256.0 & 21 \\
210 & 1042926292644833024 & 11857 & 8.56 & 0.96 & - & - & 16.9800 & 13.4900 & 0.889 & 415.0 & 21 \\
211 & 647899806626643200 & 11366 & 8.33 & 0.81 & - & - & 17.0910 & 11.1420 & 1.095 & 563.0 & 21 \\
212 & 642549544391197440 & 11017 & 7.99 & 0.60 & - & - & 16.4910 & 11.9950 & 1.347 & 783.0 & 21 \\
213 & 836410319296579712 & 10895 & 7.96 & 0.58 & - & - & 16.5180 & 11.6870 & 1.395 & 880.0 & 21 \\
214 & 1682022481467013504 & 12245 & 8.25 & 0.76 & - & - & 16.8390 & 10.9980 & 1.122 & 102.0 & 21 \\
215 & 3626525219143701120 & 11313 & 7.90 & 0.55 & - & - & 16.5110 & 10.5830 & 1.502 & 258.0 & 21 \\
216 & 1456920737222542208 & 11643 & 8.04 & 0.63 & - & - & 16.0820 & 14.1570 & 1.303 & 195.0 & 21 \\
217 & 1587611884756030720 & 11226 & 8.07 & 0.64 & - & - & 16.4564 & 12.5490 & 1.332 & 814.0 & 21 \\
218 & 4491980748701631616 & 11585 & 7.96 & 0.58 & - & - & 16.1740 & 12.9160 & 1.377 & 261.0 & 21 \\
219 & 4503347770490390016 & 10565 & 7.93 & 0.56 & - & - & 16.6310 & 11.5000 & 1.379 & 857.0 & 21 \\
220 & 2159171323461157120 & 11693 & 8.40 & 0.85 & - & - & 17.2640 & 10.5930 & 1.013 & 370.0 & 21 \\
221 & 4539136259802013952 & 10793 & 8.04 & 0.63 & - & - & 15.0360 & 24.8680 & 1.273 & 968.0 & 21 \\
222 & 2127591833389528064 & 10794 & 8.05 & 0.63 & - & - & 16.9120 & 10.5420 & 1.277 & 844.0 & 21 \\
223 & 4250461749665556224 & 11684 & 8.11 & 0.67 & - & - & 16.4850 & 12.1540 & 1.223 & 206.0 & 21 \\
224 & 4217910669267424512 & 11002 & 7.99 & 0.60 & - & - & 16.7130 & 10.8600 & 1.337 & 497.0 & 21 \\
225 & 1980205739970324224 & 10959 & 8.41 & 0.86 & - & - & 17.0820 & 12.5550 & 0.974 & 1286.0 & 21 \\
226 & 2844933221011789952 & 10798 & 7.89 & 0.54 & - & - & 16.2920 & 12.5980 & 1.432 & 277.0 & 21 \\
227 & 1913174219724912128 & 11394 & 7.98 & 0.59 & - & - & 16.2650 & 12.7370 & 1.330 & 363.0 & 21 \\
228 & 2826770319713589888 & 11185 & 7.92 & 0.56 & - & - & 16.4660 & 11.2670 & 1.434 & 1161.0 & 21 \\
229 & 2867203584218146944 & 11164 & 8.32 & 0.81 & - & - & 17.0740 & 11.5340 & 1.044 & 545.0 & 21 \\
230 & 2766498012855959424 & 11993 & 8.12 & 0.68 & - & - & 16.3540 & 12.8270 & 1.258 & 252.0 & 21 \\
231 & 4981271566317702656 & 10857 & 8.00 & 0.60 & - & - & 16.4963 & 12.1092 & 1.363 & 859.99 & 6,80,145 \\
232 & 2485615050141095424 & 9705 & 7.31 & 0.29 & - & - & 18.0653 & 4.4973 & 2.235 & 1212.0 & 55 \\
233 & 2457759374023232768 & 11872 & 7.94 & 0.57 & 12096 & 0.635 & 14.2100 & 30.5370 & 1.382 & 243.64 & 6,14,37,58,59,81,82,83,84 \\
234 & 5135466183642594304 & 10836 & 7.97 & 0.58 & 11440 & 0.570 & 15.2107 & 21.6207 & 1.372 & 642.55 & 14,22,56 \\
235 & 325647757572427776 & 11920 & 7.39 & 0.38 & - & - & 18.2589 & 9.2517 & 0.732 & 197.37 & 48 \\
236 & 2486317500631962496 & 11100 & 7.81 & 0.50 & 11568 & 0.570 & 17.9738 & 5.2900 & 1.654 & 312.55 & 6,23 \\
237 & 18949090767841024 & 10830 & 8.03 & 0.62 & - & - & 16.4827 & 12.5137 & 1.327 & 1283.7 & 48 \\
238 & 3266140045253880576 & 10790 & 7.92 & 0.56 & - & - & 17.8852 & 6.2098 & 1.488 & 763.36 & 22,23,55 \\
239 & 3263351198434024576 & 10288 & 7.87 & 0.53 & - & - & 18.2108 & 5.5103 & 1.492 & 803.31 & 22,23 \\
240 & 4835794942327992704 & 11337 & 8.01 & 0.61 & 11504 & 0.570 & 15.0418 & 22.5864 & 1.373 & 558.1 & 14,85,86,87,145,147,148 \\
241 & 5114767980330242688 & 10991 & 7.98 & 0.59 & - & - & 16.0056 & 14.6806 & 1.365 & 543.8 & 48 \\
242 & 177368203568437120 & 11015 & 7.86 & 0.53 & 11816 & 0.770 & 15.6661 & 16.0504 & 1.460 & 907.22 & 14,88,89 \\
243 & 279512833789658752 & 11227 & 7.98 & 0.59 & 11744 & 0.632 & 15.9733 & 14.6603 & 1.344 & 705.13 & 14,90,91 \\
244 & 3238868171156736768 & 11487 & 8.08 & 0.66 & 12256 & 0.660 & 15.3757 & 20.1271 & 1.255 & 485.19 & 6,14,37,92,93,94,95,96 \\
245 & 3446909137068558464 & 11727 & 8.04 & 0.63 & 12064 & 0.593 & 15.5794 & 17.5116 & 1.299 & 287.14 & 6,14,22,37,51,71,97,98 \\
246 & 952987242221386496 & 10516 & 8.05 & 0.63 & - & - & 15.2767 & 22.9162 & 1.278 & 1366.4 & 20 \\
247 & 924428973878531968 & 11028 & 8.03 & 0.62 & - & - & 15.8089 & 16.6271 & 1.313 & 633.44 & 48 \\
248 & 928600654136951936 & 12150 & 8.27 & 0.77 & - & - & 19.2758 & 2.8108 & 1.475 & 388.39 & 22,23 \\
249 & 651567743057943424 & 13037 & 8.25 & 0.76 & 11104 & 0.510 & 19.0012 & 3.8737 & 1.129 & 288.71 & 26,146 \\
250 & 911639111025648512 & 11365 & 7.99 & 0.60 & 11232 & 0.609 & 15.6441 & 16.9660 & 1.348 & 491.28 & 14,99,100,101 \\
251 & 609528255272111360 & 10593 & 7.92 & 0.56 & 11512 & 0.749 & 18.3706 & 5.0191 & 1.432 & 313.08 & 26,146 \\
252 & 909913766832936576 & 11447 & 7.96 & 0.58 & - & - & 18.8781 & 3.7588 & 1.397 & 301.1 & 22,23 \\
253 & 1013314902617375104 & 11649 & 7.73 & 0.46 & - & - & 18.3897 & 3.9771 & 1.620 & 189.19 & 22,23 \\
254 & 582901519620616192 & 11314 & 7.95 & 0.58 & - & - & 16.8173 & 9.6431 & 1.401 & 326.0 & 29 \\
255 & 660485847511385984 & 11713 & 7.89 & 0.54 & 11632 & 0.663 & 18.9457 & 3.4279 & 1.472 & 201.3 & 26,146 \\
256 & 612465222627864192 & 11619 & 7.98 & 0.59 & 10816 & 0.853 & 17.6740 & 6.4549 & 1.398 & 244.6 & 26,146 \\
257 & 715357559411376128 & 11576 & 8.10 & 0.66 & 11392 & 0.660 & 14.5563 & 29.4257 & 1.242 & 936.84 & 14,22,102 \\
258 & 612200652642509952 & 10888 & 7.95 & 0.48 & 11512 & 0.749 & 19.4093 & 2.7153 & 1.648 & 805.81 & 26,103,146 \\
259 & 3842810971782221056 & 11376 & 8.07 & 0.64 & - & - & 17.8421 & 6.4916 & 1.277 & 601.37 & 23 \\
260 & 578824942821404544 & 11217 & 8.11 & 0.67 & - & - & 18.5235 & 4.9743 & 1.255 & 399.65 & 29 \\
261 & 590669363112793600 & 11359 & 7.97 & 0.59 & - & - & 18.1809 & 5.2086 & 1.399 & 262.12 & 29 \\
262 & 1025369256613064192 & 11009 & 8.06 & 0.64 & - & - & 17.4672 & 7.9325 & 1.330 & 619.64 & 22,23 \\
263 & 3834020410677227904 & 11430 & 8.36 & 0.83 & - & - & 18.8301 & 3.7313 & 1.458 & 435.23 & 23 \\
264 & 615037976757661696 & 11947 & 8.00 & 0.61 & - & - & 16.5794 & 10.6021 & 1.354 & 249.95 & 22,23 \\
265 & 626319721972908288 & 10948 & 7.99 & 0.60 & - & - & 16.3546 & 12.7590 & 1.350 & 1078.01 & 22,23 \\
266 & 3834519765050176640 & 11702 & 7.85 & 0.52 & - & - & 16.7892 & 8.9069 & 1.499 & 223.13 & 22,23 \\
267 & 1048501022756210432 & 11332 & 8.00 & 0.60 & - & - & 18.1361 & 5.4151 & 1.404 & 466.13 & 22,23 \\
268 & 3859871342090899328 & 11806 & 8.03 & 0.62 & - & - & 15.6659 & 16.5554 & 1.297 & 244.83 & 23 \\
269 & 779588657882981120 & 11005 & 7.96 & 0.58 & - & - & 16.1299 & 13.6912 & 1.419 & 849.67 & 30 \\
270 & 3804750586512019840 & 11314 & 7.94 & 0.57 & - & - & 17.6110 & 6.6610 & 1.490 & 826.53 & 22,23 \\
271 & 3805226228370453504 & 10703 & 7.97 & 0.59 & - & - & 18.4054 & 5.0840 & 1.447 & 912.04 & 29 \\
272 & 3813460436591500800 & 10706 & 8.01 & 0.60 & 11904 & 0.870 & 18.1167 & 5.8887 & 1.397 & 814.61 & 22,23,26,146 \\
273 & 3812725275629199360 & 11023 & 7.88 & 0.53 & - & - & 18.0952 & 5.2659 & 1.526 & 260.69 & 22,23 \\
274 & 3541237717085787008 & 11505 & 8.00 & 0.60 & - & - & 16.3940 & 11.9447 & 1.385 & 287.34 & 18 \\
275 & 769375702394078208 & 11561 & 8.03 & 0.62 & 11240 & 0.632 & 16.5968 & 11.0528 & 1.389 & 340.81 & 14,99,104 \\
276 & 3897445571422443904 & 10839 & 8.01 & 0.61 & 11336 & 0.570 & 14.9625 & 24.8024 & 1.315 & 1009.07 & 14,18,26,146 \\
277 & 3898180418851205888 & 10647 & 7.98 & 0.59 & - & - & 17.5531 & 7.5071 & 1.381 & 810.29 & 23 \\
278 & 1717969674107797760 & 11409 & 8.08 & 0.65 & 11184 & 0.660 & 16.0183 & 15.1997 & 1.258 & 722.47 & 14,22,105,106 \\
279 & 3905320235045590400 & 10899 & 7.90 & 0.55 & - & - & 18.5422 & 4.5029 & 1.478 & 718.25 & 29 \\
280 & 1465305681615674624 & 11925 & 8.26 & 0.77 & - & - & 18.0926 & 6.3157 & 1.167 & 364.6 & 20 \\
281 & 3690690583141563520 & 10707 & 7.98 & 0.59 & - & - & 19.1167 & 3.7124 & 1.445 & 909.67 & 29 \\
282 & 3690323316193465344 & 11017 & 7.91 & 0.55 & 11584 & 0.632 & 16.3444 & 12.0698 & 1.485 & 741.65 & 6,14,29,37 \\
283 & 1474090607723007104 & 11026 & 8.03 & 0.62 & 11576 & 0.705 & 15.3223 & 20.8288 & 1.309 & 1079.8 & 14,107,120,121,122,123 \\
284 & 3684922377638797312 & 10462 & 7.40 & 0.32 & - & - & 17.6873 & 5.0052 & 2.094 & 324.98 & 29 \\
285 & 3618569530962200832 & 10820 & 8.00 & 0.60 & 11816 & 0.681 & 16.6728 & 11.2791 & 1.376 & 400.69 & 26,146 \\
286 & 6191268135405788544 & 10591 & 8.00 & 0.60 & - & - & 16.0736 & 15.1149 & 1.349 & 1088.05 & 18 \\
287 & 2101020107679847552 & 11196 & 7.88 & 0.54 & 11848 & 0.675 & 18.2417 & 4.8473 & 1.443 & 339.19 & 26,28,146 \\
288 & 2105615585247026688 & 11268 & 8.08 & 0.65 & 11624 & 0.510 & 18.1241 & 5.7989 & 1.268 & 294.59 & 26,28,146 \\
289 & 2130795019997375360 & 11936 & 8.12 & 0.68 & 11584 & 0.705 & 19.0071 & 3.9252 & 1.186 & 784.59 & 26,28,146 \\
290 & 2133056165659732480 & 11649 & 8.02 & 0.62 & 11064 & 0.729 & 18.0637 & 5.4976 & 1.359 & 249.54 & 26,124,146 \\
291 & 3800979880104156032 & 12520 & 8.19 & 0.72 & 11752 & 0.710 & 16.8945 & 7.6345 & 1.561 & 223.4 & 26,146 \\
292 & 3897443612917355008 & 12578 & 8.15 & 0.69 & 10672 & 0.858 & 17.6394 & 6.9136 & 1.216 & 266.23 & 26,146 \\
293 & 2615269564367501952 & 10656 & 8.00 & 0.60 & 11608 & 0.753 & 17.3605 & 8.4222 & 1.375 & 1192.5 & 26,146 \\
294 & 38532728593335296 & 10135 & 7.51 & 0.36 & 11336 & 0.599 & 17.6221 & 5.9430 & 2.077 & 833.36 & 26,146 \\
295 & 659723370557599872 & 11936 & 8.35 & 0.83 & - & - & 18.8701 & 4.8203 & 1.056 & 161.79 & 26,146 \\
296 & 603166652792073088 & 11687 & 8.19 & 0.72 & 11728 & 0.662 & 18.2364 & 5.7467 & 1.181 & 301.36 & 26,146 \\
297 & 6769034244236810368 & 10217 & 7.92 & 0.56 & 11624 & 0.772 & 17.7248 & 7.1657 & 1.416 & 119.34 & 26,146 \\
298 & 2534749789540201728 & 11250 & 8.45 & 0.89 & 11360 & 0.672 & 17.9344 & - & - & 642.64 & 26,146 \\
299 & 2538281386529032192 & 11645 & 8.01 & 0.61 & 11752 & 0.815 & 16.3282 & 12.2592 & 1.339 & 205.92 & 26,146 \\
300 & 2551666050811485440 & 11709 & 7.98 & 0.59 & 10600 & 0.840 & 17.6621 & 6.4725 & 1.462 & 209.31 & 26,146 \\
301 & 2555914937403311232 & 11059 & 8.02 & 0.61 & 12200 & 0.620 & 17.9110 & 6.2731 & 1.438 & 989.33 & 26,146 \\
302 & 1561515526026874624 & 11493 & 7.90 & 0.55 & 12208 & 0.609 & 15.8296 & 14.4025 & 1.525 & 260.05 & 14,37,75 \\
303 & 1671329451713079808 & 11731 & 7.93 & 0.57 & 11496 & 0.609 & 15.5423 & 16.5889 & 1.460 & 178.13 & 14,108,109 \\
304 & 3663900436870097664 & 11435 & 7.87 & 0.53 & - & - & 16.4282 & 10.8699 & 1.536 & 222.61 & 23 \\
305 & 6301064404482150912 & 11691 & 8.06 & 0.64 & 11712 & 0.632 & 15.7160 & 16.8216 & 1.306 & 811.42 & 6,14,22,37,79,110,111,112,113 \\
306 & 3671324781762360448 & 11294 & 8.04 & 0.63 & - & - & 17.9724 & 6.0709 & 1.384 & 876.94 & 29 \\
307 & 3666095916777028608 & 11163 & 7.99 & 0.59 & - & - & 18.1513 & 5.4855 & 1.435 & 836.18 & 22,23 \\
308 & 1176717792385803136 & 12222 & 8.05 & 0.63 & 11632 & 0.632 & 14.3435 & 29.9941 & 1.293 & 353.92 & 14,34,114,115,116 \\
309 & 3655250883836759808 & 11068 & 8.00 & 0.60 & - & - & 18.7301 & 4.3305 & 1.408 & 1016.44 & 23 \\
310 & 4419519835755847936 & 11388 & 7.87 & 0.53 & - & - & 18.6953 & 3.9290 & 1.599 & 524.48 & 23,29 \\
311 & 4415701304886958848 & 11868 & 8.05 & 0.64 & - & - & 15.8146 & 15.5399 & 1.310 & 710.88 & 22,23,117 \\
312 & 1697790612080870144 & 12012 & 8.34 & 0.82 & 12496 & 0.770 & 16.5405 & 13.4467 & 1.078 & 112.5 & 14,18 \\
313 & 1641326979142898048 & 11239 & 8.01 & 0.61 & 11760 & 0.609 & 15.5759 & 17.9189 & 1.367 & 636.12 & 14,118,147 \\
314 & 1372458109403442432 & 10983 & 7.96 & 0.58 & 11216 & 0.705 & 14.4081 & 30.4668 & 1.356 & 828.7 & 6,14,64,89 \\
315 & 1204873051062881408 & 10256 & 7.01 & 0.22 & - & - & 17.3229 & 4.7496 & 2.602 & 1928.5 & 20 \\
316 & 1382407895064799616 & 10982 & 7.91 & 0.55 & - & - & 18.4106 & 4.7079 & 1.547 & 776.66 & 22,23 \\
317 & 5923100131319630976 & 10759 & 7.93 & 0.56 & 11392 & 0.609 & 15.5821 & 17.8586 & 1.381 & 1082.84 & 14,37,119 \\
318 & 1433262339369791872 & 11503 & 7.92 & 0.56 & - & - & 17.5918 & 6.5714 & 1.492 & 282.7 & 23 \\
319 & 1434744751625887616 & 10617 & 7.89 & 0.54 & - & - & 18.7424 & 4.1626 & 1.520 & 1210.27 & 22,23 \\
320 & 2158083013107793792 & 11300 & 7.58 & 0.39 & 11232 & 0.570 & 16.1487 & 10.4044 & 1.809 & 292.77 & 14,18,125 \\
321 & 2126916119073072384 & 9954 & 7.61 & 0.40 & - & - & 18.4624 & 4.5104 & 1.630 & 310.9 & 28 \\
322 & 2053023676347758976 & 11055 & 8.10 & 0.66 & - & - & 19.4114 & 3.3901 & 1.224 & 723.6 & 28 \\
323 & 2025389380082340992 & 11657 & 8.01 & 0.61 & 11720 & 0.660 & 13.0613 & 54.7767 & 1.308 & 249.2 & 14,90,127,128,129 \\
324 & 2080034588234814464 & 11440 & 8.16 & 0.70 & - & - & 18.4719 & 5.1661 & 1.192 & 766.0 & 28 \\
325 & 2078037050483671552 & 11231 & 7.94 & 0.57 & - & - & 16.7623 & 9.9248 & 1.386 & 322.0 & 28 \\
326 & 2079732192477615360 & 10645 & 7.80 & 0.49 & - & - & 17.1646 & 8.1476 & 1.496 & 255.9 & 28 \\
327 & 4248931504366754304 & 10362 & 7.99 & 0.60 & - & - & 16.3832 & 13.5427 & 1.325 & 1350.4 & 48 \\
328 & 1840865211187303424 & 11469 & 8.03 & 0.62 & - & - & 15.9849 & 14.7846 & 1.287 & 800.0 & 130 \\
329 & 2173871656498100096 & 10896 & 7.86 & 0.52 & 10952 & 0.593 & 16.8547 & 9.3008 & 1.533 & 959.01 & 14,131 \\
330 & 2680093711924796928 & 9891 & 7.41 & 0.33 & - & - & 18.3755 & 4.1238 & 1.965 & 1346.81 & 55 \\
331 & 2653002398251615872 & 10722 & 8.15 & 0.69 & - & - & 18.8843 & 4.5762 & 1.247 & 637.54 & 55 \\
332 & 2719012579552367488 & 11371 & 7.98 & 0.59 & 12424 & 0.593 & 15.7612 & 15.9821 & 1.348 & 290.65 & 14,51,71 \\
333 & 2842650153836732928 & 11013 & 7.96 & 0.58 & 11208 & 0.525 & 15.3516 & 19.6918 & 1.359 & 708.07 & 14,132,133,134,135 \\
334 & 2660358032257156736 & 11117 & 8.00 & 0.60 & 11472 & 0.593 & 13.0624 & 57.0620 & 1.323 & 612.82 & 6,14,37,88,136,137,138,139,152 \\
335 & 2338349628107880192 & 11195 & 8.02 & 0.62 & 11696 & 0.770 & 15.3551 & 20.0211 & 1.320 & 846.51 & 14,78,140 \\
336 & 3892640293292355328 & 10400 & 7.39 & 0.32 & 10920 & 0.338 & 18.4987 & 1.5829 & 4.652 & 1141.7 & 142,143 \\
337 & 3305381267356525824 & 9807 & 7.76 & 0.47 & 10952 & 0.575 & 18.4530 & 5.2370 & 1.776 & 715.1 & 146 \\
338 & 1999127510441929600 & 11005 & 8.04 & 0.63 & 12112 & 0.760 & 16.8802 & 10.2350 & 1.296 & 857.88 & 149 \\
339 & 2345323551189913600 & 12694 & 9.31 & 1.30 & 13024 & 1.310 & 19.0352 & 10.0396 & 0.431 & 215.58 & 150 \\
\\
\hline
\end{longtable}
\end{landscape} % End Horizontal Page
\twocolumn

\end{document}